\documentclass[11pt]{article}  

\usepackage[T1]{fontenc}
\usepackage[latin1]{inputenc} 
\usepackage{graphicx}  
\usepackage{fancyhdr,lastpage}
\usepackage[NewCommands]{ragged2e}
\usepackage{amsmath}
\usepackage{latexsym}
\usepackage{url}
\usepackage{hyperref}
\usepackage[dvips]{color}
\usepackage[sort&compress,numbers]{natbib}
\usepackage{pstricks,pst-node,pst-coil,pst-plot}

\begin{document}

\renewcommand{\textfraction}{0.05}
\renewcommand{\topfraction}{0.95}
\renewcommand{\bottomfraction}{0.95}
\renewcommand{\floatpagefraction}{0.95}
\renewcommand{\baselinestretch}{1.095} 
\setlength{\parindent}{0pt}  
\setlength{\parskip}{5pt plus 2pt minus 1pt}
 \renewcommand{\footnoterule}{\vspace{0.5cm}
 \rule{2.5in}{0.4pt} \vspace{0.3cm}}

\newcounter{myfn}[page]
\renewcommand{\thefootnote}{\fnsymbol{footnote}}
\newcommand{\myfootnote}[1]{
  \setcounter{footnote}{\value{myfn}}
  \footnote{#1}\stepcounter{myfn}} \newcommand{\fn}[1]{\myfootnote{
    #1}}
 \newcommand{\mean}[1]{\left\langle #1 \right\rangle}
 \newcommand{\abs}[1]{\left| #1 \right|}
 \newcommand{\la}{\langle}
 \newcommand{\ra}{\rangle}
 \newcommand{\RA}{\Rightarrow}
 \newcommand{\tet}{\vartheta}
 \newcommand{\eps}{\varepsilon}
 \newcommand{\bbox}[1]{\mbox{\boldmath $#1$}}
 \newcommand{\ul}[1]{\underline{#1}}
 \newcommand{\ol}[1]{\overline{#1}}
 \newcommand{\non}{\nonumber \\}
 \newcommand{\no}{\nonumber}
 \newcommand{\eqn}[1]{eq. (\ref{#1})}
 \newcommand{\Eqn}[1]{Eq. (\ref{#1})}
 \newcommand{\eqs}[2]{eqs. (\ref{#1}), (\ref{#2})}
 \newcommand{\pics}[2]{Figs. \ref{#1}, \ref{#2}}
 \newcommand{\pic}[1]{Fig. \ref{#1}}
 \newcommand{\sect}[1]{Sect. \ref{#1}}
 \newcommand{\name}[1]{{\rm #1}}

\newcommand{\rez}{\hbox{\rm I$\!$R}}
\newcommand{\naz}{\hbox{\rm I$\!$N}}

\newcommand{\rb}[1]{\raisebox{-1 ex}{#1}}
\newcommand{\av}[1]{\left< #1 \right>}
\newcommand{\abst}[0]{\rule[-1.5 ex]{0 ex}{4 ex}}
\newcommand{\fref}[1]{Fig.~\ref{#1}}
\newcommand{\tref}[1]{Tab.~\ref{#1}}
\newcommand{\eref}[1]{Eq.~(\ref{#1})}
\newcommand{\sref}[1]{Section~\ref{#1}}
\newcommand{\aref}[1]{Appendix~\ref{#1}}

 \newcommand{\vol}[1]{{\bf #1}}
 \newcommand{\et}{{\it et al.}}
 \newcommand{\D}{\displaystyle}
 \newcommand{\T}{\textstyle}
 \newcommand{\SC}{\scriptstyle}
 \newcommand{\SSC}{\scriptscriptstyle}
 \renewcommand{\textfraction}{0.05}
 \renewcommand{\topfraction}{0.95}
 \renewcommand{\bottomfraction}{0.95}
 \renewcommand{\floatpagefraction}{0.95}

\vspace*{0.2cm}
 \begin{center}
   \textbf{\Large Modeling Vortex Swarming In Daphnia}\\[5mm]
{\large \bf  Robert Mach, Frank Schweitzer$^{\star}$} 

\begin{quote}
\begin{itemize}
\item[] \emph{Chair of Systems Design, ETH Zurich, Kreuzplatz 5,
    8032 Zurich, Switzerland}
\item[] $^{\star}$ corresponding author, email:
    \url{fschweitzer@ethz.ch}
\end{itemize}
\end{quote}

\end{center}

\begin{abstract}
  Based on experimental observations in \textit{Daphnia}, we introduce an
  agent-based model for the motion of single and swarms of animals.  Each
  agent is described by a stochastic equation that also considers the
  conditions for active biological motion. An environmental potential
  further reflects local conditions for \textit{Daphnia}, such as
  attraction to light sources.  This model is sufficient to describe the
  observed cycling behavior of single \textit{Daphnia}. To simulate
  vortex swarming of many \textit{Daphnia}, i.e. the collective rotation
  of the swarm in one direction, we extend the model by considering
  avoidance of collisions. Two different ansatzes to model such a
  behavior are developed and compared. By means of computer simulations
  of a multi-agent system we show that local avoidance - as a special
  form of asymmetric repulsion between animals - leads to the emergence
  of a vortex swarm.  The transition from uncorrelated rotation of single
  agents to the vortex swarming as a function of the swarm size is
  investigated.  Eventually, some evidence of avoidance behavior in
  \textit{Daphnia} is provided by comparing experimental and simulation
  results for two animals. 

\end{abstract}

\section{Introduction}

Swarming is a prominent example of complex behavior in biological
systems.  This form of \emph{collective motion} may emerge from the
interplay of individual behavior and \emph{local interactions} of a large
number of individuals (agents). Swarms (also called herds, flocks,
schools) can often be observed in certain mammals, fish, insects, and
birds for various benefits, such as enhanced feeding and mating as well
as more successful predator avoidance \citep[see
e.g.][]{parrish-hamner-ed-97,parrish02:_self_fish_school,
  Parrish:1999,Okubo:2002,couzin01:social_organisation_fish_schools,huth94,huth92}.
This has been reported for several prey animals, e.g. in planktivore fish
\citep{Partridge:1982,Hall:1986}, in some species of birds
\citep{Caraco:1980}, as well as in zooplankton
\citep{Jakobsen:1994,Kvam:1995}.

Detailed experimental investigations on swarming, however, are rare,
either because of the size of the animals or because well defined
conditions for experiments are difficult to realize.  Earlier chance
observations were reported for horizontally circling zooplankton in
the field \citep{Lobel:1986}. They triggered further experiments with
\textit{Daphnia} relevant for the current paper (see also Sects.
\ref{sec:experimental-methods} and \ref{sec:gath-empir-evid}). It has
been shown that under certain circumstances single \textit{Daphnia}
circle horizontally around a vertical artificial light shaft to which
they are attracted.  For high \textit{Daphnia} densities a swarm
emerges, where all \textit{Daphnia} circle in the same randomly chosen
direction~\citep{ordemann02:_vortex_swarm_zoopl_daphn,ordemann03:_pattern_formation,ordemann03:_motion_daphn_light_field}.

The physical, biological, and chemical reasons for swarming in
\textit{Daphnia} in particular and in prey animals in general are not
completely understood.  Biological considerations suggest that circling
is the least energy consuming motion for permanently moving animals to
stay as a group at a certain local position without frequently bumping
into each other.  Recent studies concentrated on the response of
individuals as well as groups to various external influences, such as
available food, food gradients and predator threat
\citep{jakobsen87:_behav_daphn,larsson95:_food_daphn,kleiven96:_direc_daphn,larsson97:_ideal_daphn,jensen00:_gregar_daphn}.
In particular, \citet{oeien04:_daphn_dynam_based_kinet_theor} used
methods from plasma kinetic theory to derive macroscopic equations --
so-called fluid-dynamic equations -- for the density of Daphnicle
(Daphnia-like particle) swarms as a function of food-concentration, food
saturation of Daphnia and a threat field of predators. The advantage of
this approach is, that these equations can be solved analytically by
approximations in linear space. However, it is difficult to compare these
results for swarming to rotating real \emph{Daphnia} swarms, as we are
interested in.

While \citet{oeien04:_daphn_dynam_based_kinet_theor} concentrates his
description on the macroscopic level (density ond velocity destributions
of the swarm), we aim at understanding swarming from a ``microscopic''
approach. That means, we derive equations of motion for individual
entities -- so-called agents -- and investigate the collective motion by
means of a multi-agent system. This allows us to understand the behavior
at the system level from the interactions of the entities comprising
the system.

Individual-based or \emph{agent-based} modeling has turned out to be a
very useful tool for modeling biological phenomena at various levels of
organization \citep{flierl-et-99, gruenbaum-okubo-94, deutsch-99,
  kunz03:_artif_fish_school}. Thus, recently different computer
architectures have been developed to simulate the collective behavior of
interacting agents in distributed artificial intelligence (see e.g.
\verb+http://www.swarm.org/+). However, due to their rather complex
simulation facilities many of these simulation tools lack the possibility
to investigate systematically and in depth the influence of specific
interactions and parameters. Instead of incorporating only as much detail
as is \emph{necessary} to produce a certain emergent behavior, they put
in as much detail \emph{as possible}, and thus reduce the chance to
understand \emph{how} emergent behavior occurs and \emph{what} it depends
on.  Therefore, in this paper, we follow a different multi-agent
approach, based on \emph{Brownian agents}
\citep{schweitzer03:_brown_agent_activ_partic} (see
\sect{sec:agent-model-biol}), that -- in addition to its computational
suitability -- can be also investigated by means of analytical methods
from statistical physics and mathematics.

The objective of the present study is to investigate the requirements on
the microscopic level that lead to the formation of a vortex swarm on the
macroscopic level. A swarm is called \emph{vortex} swarm if animals
\emph{cycle} around an imaginary axis in the same rotational direction.
On the global level (i.e. in three dimensions) one observes the emergence
of a cone.  As the main (cycling) motion takes place in the horizontal
plane, we have restricted both the experimental observations and our
simulation to two dimensions for simplicity.  To find out more about
physical reasons for vortex swarming, we first summarize some
experimental observations previously reported, and then
set-up a ``minimalistic'' multi-agent model to test some biologically
relevant assumptions that may lead to the observed swarming behavior.

\section{Experimental Observations on Daphnia Motion}
\label{sec:experimental-methods}

In this section, we summarize some biological facts about \emph{Daphnia}
animals and their collective movement which have been reported in the
literature. This information shall be used to motivate our swarming model
in the following sections. 

The water flea or \textit{Daphnia} (see \pic{fig:daphnia}\footnote{ The
  photograph is reprinted with the permission of Stephen Durr,
  \url{http://www.btinternet.com/~stephen.durr/photographthree.html}.
  For more detailed information about Daphnia, see e.g.
  \url{http://www.lander.edu/rsfox/310DaphniaLab.html},
  \url{http://ebiomedia.com/gall/classics/Daphnia/} }) is a member of the
\emph{crustacea} and are found in most fresh water ponds.  Their body is
enclosed within a carapace and their length is about 1mm to 3mm.
\textit{Daphnia} swims with a jerky motion through the water as the
powerful 2nd antennae are thrust downward.
\begin{figure}[htbp]
\centerline{
\includegraphics[width=7cm]{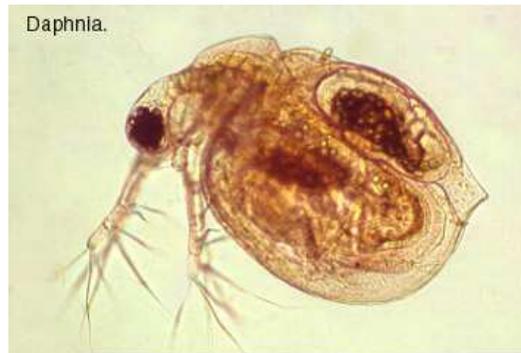}
}
\caption{Daphnia, Courtesy of Stephen Durr.}
\label{fig:daphnia}
\end{figure}

Ecologically, \textit{Daphnia} is extremely important in the food chains
of ponds and lakes. So a systematic investigation of their individual and
collective behavior is of great interest.  Ordemann \emph{et al.}
\citep{ordemann02:_vortex_swarm_zoopl_daphn,ordemann03:_pattern_formation,ordemann03:_motion_daphn_light_field}
have experimentally investigated the motion of \textit{Daphnia} close to
a vertical light shaft in both \emph{low density} and \emph{high density}
\textit{Daphnia} swarms, as we summerize in the following.

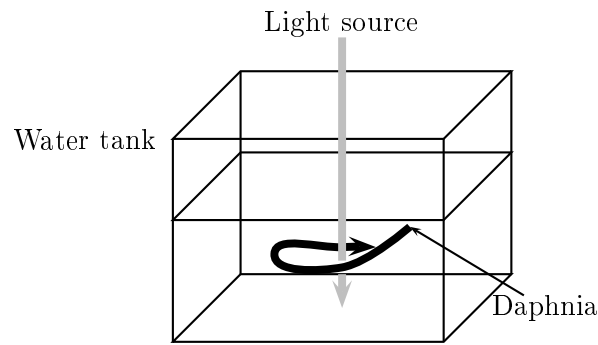
\begin{figure}[htbp]
\begin{center}
{\psset{unit=0.9cm}
\begin{pspicture}(-3,-1)(3,5)
\pspolygon(-2.5, 2.5)(-1.5, 3.5)(2.5, 3.5)(1.5, 2.5)
\pspolygon(-2.5,-0.5)(-1.5,0.5)(2.5,0.5)(1.5,-0.5)
\pspolygon(-2.5,1.3)(-1.5,2.3)(2.5,2.3)(1.5,1.3)
\psline(-2.5,-0.5)(-2.5,2.5)
\psline(-1.5,0.5)(-1.5,3.5)
\psline(2.5,0.5)(2.5,3.5)
\psline(1.5,-0.5)(1.5,2.5)

\pscurve[linewidth=3pt]{->}(1,1.2)(0,0.6)(-1,0.8)(0,0.9)(0.5,0.9)
\psline[linecolor=lightgray,linewidth=3pt](0,4)(0,0.7)
\psline[linecolor=lightgray,linewidth=3pt]{->}(0,0.5)(0,0)
\pnode(1,1.2){Daphnia}
\rput(3,0){\rnode{name}{Daphnia}}
\ncline{->}{name}{Daphnia}
\rput(0,4.2){Light source}
\rput(-3.8,2.5){Water tank}
\end{pspicture}}
\vspace*{-10pt}
\end{center}
\caption[]{Sketch of the \textit{Daphnia} experiments carried out by  Ordemann
  \emph{et al.}
  \citep{ordemann02:_vortex_swarm_zoopl_daphn,ordemann03:_pattern_formation,ordemann03:_motion_daphn_light_field}.
  \label{experiment}}
\end{figure}

Using the experimental outline shown in \pic{experiment}, it was found
that a \emph{single} \textit{Daphnia} is attracted to the light and
starts to \emph{cycle} (i.e. rotate) around the artificial light source,
keeping its cycling direction for quite a while. In repeated experiments,
however, the cycling direction may change to the opposite, which leads to
the conclusion that single \textit{Daphnia}, while rotating around the
light beam, do \emph{not} have a preferred direction of motion. 

The same behavior has been observed for \textit{Daphnia} swarms of
sufficiently low animal density, where individual animals close to the
light shaft cycle in both directions around the light shaft, frequently
changing their cycling direction.  

Interestingly, the situation changes if instead of single or few
\textit{Daphnia} a large number of animals is put in the water tank.  In
this case, the \textit{Daphnia} start again with their cycling motion,
but then \emph{all} tend to move into the \emph{same} direction of
motion.  From a physical perspective, a symmetry break is observed, i.e.,
the symmetry between the two possible cycling directions (left, right
rotation) is clearly broken toward \emph{one} of the possibilities (left
\emph{or} right rotation).  Both of these possibilities have the same
chance to occur, but only one of them is eventually realized.  The vortex
formation as well as the symmetry break in the cycling direction are
clearly self-organized phenomena that result from the \emph{collective}
interaction of \emph{many} animals.  In order to understand this in more
detail, we derive a multi-agent model in the following.

\section{Agent model of biological motion in an environmental potential}
\label{sec:agent-model-biol}

Our modeling approach is based on active Brownian particles
 or \emph{Brownian agents},
respectively \citep{schweitzer03:_brown_agent_activ_partic}. Each of
these agents is described by three state variables: spatial position
$\bbox{r}_{i}$, velocity $\bbox{v}_{i}$ and internal energy depot
$e_{i}$. The first two state variables describe the \emph{movement} of
the agent and can be observed from the outside. The agent's energy depot,
however, is an \emph{internal} variable that considers the take-up of
energy from the environment, the storage of energy and conversion of
stored energy into energy of motion \citep{fs-eb-tilch-98-let}. Provided
a supercritical supply of energy from the environment, the Brownian agent
is capable of \emph{active movement}, e.g. in a preferred direction. The
term ``Brownian'' refers to the fact that the agents may still be subject
to fluctuations that are described by a stochastic force.

The model of Brownian agents was widely discussed in different
publications
\citep{fs-eb-tilch-98-let,eb-fs-tilch-98,tilch-fs-eb-99,ebeling02:_self_organ_activ_brown_dynam_biolog_applic,erdmann-et-00}.
Therefore, only the basic dynamics are summarized here. For the external
variables $\bbox{r}_{i}$ and $\bbox{v}_{i}$, we find the equations of
motion in the form of a generalized Langevin equation:
\begin{equation}
\label{model-2}
\frac{d}{dt}\bbox{r}_{i} =  \bbox{v}_{i} \;;\quad
\frac{d}{dt}\bbox{v}_{i}  =  - \gamma(v_{i}^{2})\,\bbox{v}_{i}
-\left.\bbox{\nabla} U(\bbox{r})\right|_{r_{i}}
+ \sqrt{2D} \bbox{\xi}_{i}(t)
\end{equation}
Here, for the mass $m=1$ is used. Causes for the change of the variables
are summarized on the right-hand side of the equations.  The change of
the agent's position, $\bbox{r}_{i}$ is caused by the movement of the
agent, described by the velocity $\bbox{v}_{i}$, that in turn can be
changed by three different forces, explained in the following. 
The first term, $\gamma(v_{i}^{2})$, is a \emph{non-linear friction
  function} \citep{erdmann-et-00,
  ebeling02:_self_organ_activ_brown_dynam_biolog_applic}:
\begin{equation}
  \label{gamma-v2}
 \gamma(v_{i}^{2})=\gamma_{0}- d_{2}\,e_{i}(t) 
=\gamma_{0}- \frac{d_{2}\,q_{0}}{c+d_{2}v_{i}^{2}}
\end{equation}
which considers the active motion of the agent.  $\gamma_{0}$ is the
friction coefficient known from passive Brownian motion, whereas the
other terms describe the influence of the internal energy depot
$e_{i}(t)$, which mainly compensates this friction. Assuming that the
internal energy depot relaxes very fast into a quasi-stationary
equilibrium (adiabatic approximation), we derived an expression for the
quasi-stationary energy depot dependent on the characteristic parameters
describing its dynamics: $q_{0}$ is the influx of energy into the
internal depot, which is assumed as constant here.  $c$ describes the
loss of energy due to internal dissipation (metabolism), whereas $d_{2}$
describes the conversion rate of internal energy into kinetic energy. The
nonlinear friction function has a zero for
\begin{equation}
  \label{v0}
  v_{0}^{2}=\frac{q_{0}}{\gamma_{0}}-\frac{c}{d_{2}}
\end{equation}
\emph{Active motion}, i.e. $\abs{v_{0}}>0$ becomes possible only for a
certain supercritical take-up of energy from the environment,
$q_{0}>c\gamma_{0}/d_{2}$. 

The second term in \eqn{model-2}, $\bbox{\nabla} U(\bbox{r})$, is used to
describe the influences of the environment.  The actual motion of the
agent is a \emph{compromise} between its active motion -- which
eventually would lead it everywhere, as long as internal energy is
provided -- and the environmental conditions which set some restrictions
on this motion.  The experiments described above have used a vertical
beam of light that causes an attractive force on the Daphnia, which tend
to cycle around it. In order to cope with this, we may choose the very
simple assumption of an \emph{environmental potential} of the form
\begin{equation}
\label{eq:external_potential}
U(\bbox{r})=\frac{a}{2}\bbox{r}^2
\end{equation}
which generates an attractive force $\bbox{F}=-\bbox{\nabla}
U(\bbox{r})=-a\bbox{r}$ towards the center, $\bbox{r}=0$.

Eventually, the last term in \eqn{model-2} is a stochastic force
$\bbox{\xi}$ (assumed to be Gaussian white noise) of strength $D$, which
describes the influence of random events on the agent's motion.

The equation of motion for the Brownian agents, \eqn{model-2}, is
formulated by using two dynamical variables, $\bbox{r}_{i}(t)$,
$\bbox{v}_{i}(t)$, as originally proposed by Langevin for the motion of
Brownian particles. In this description, fluctuations in the environment
are summarized in a stochastic \emph{force} that changes the
\emph{acceleration} of the particles (according to Newton's law of
motion). In the so-called overdamped case, one can derive from this
equation the overdamped Langevin equation by assuming a quasistationary
velocity, $\dot{v}\approx 0$. This results in only one equation for
$\dot{r}$, where the stochastic term appears with a different prefactor,
$\sqrt{2D_{r}}$, $D_{r}$ being the \emph{spatial diffusion coefficient}.
While such an approximation is convenient for further theoretical
investigations, in the following we use \eqn{model-2} for our computer
simulations.

\pic{single_brownian_particle_wi_pump_wi_pot} shows computer simulations
for the active movement of a single agent, bound by an environmental
potential, \eqn{eq:external_potential}, as described by \eqn{model-2} The
result clearly indicates the \emph{cyclic motion} round the center, which
has been also observed in single Daphnia motion, as explained above.
Running the computer simulations for single agents with different initial
conditions eventually results in the same kind of cyclic motion, but with
different rotational directions, i.e. left-handed or right-handed
rotations.  Due to stochastic influences, also changes of the direction
of motion become possible.  Thus, we may conclude that our model of
Brownian agents sufficiently describes the observed behavior of
\emph{single Daphnia}.
\begin{figure}[htbp]
  \centerline{\includegraphics[width=7cm]{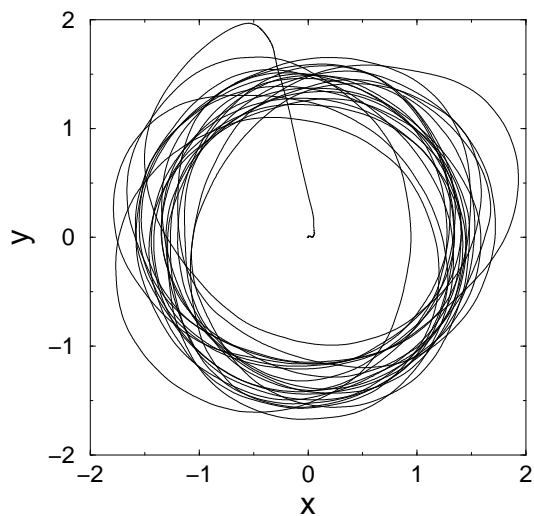}}
\caption[]{\label{single_brownian_particle_wi_pump_wi_pot}
  Trajectory of a single Brownian agent moving in an environmental
  potential, \eqn{eq:external_potential}, after $t=200$. Parameters:
  $\gamma=5.0$, $d_2=1.0$, $q_0=10.0$, $c=1.0$, $D=0.005$, $a=0.5$, i.e.
  supercritical take-up of energy, $q_{0}>c\gamma_{0}/d_{2}$. Initial
  conditions: $\{x(0),y(0)\}=\{0,0\}$, $\{v_{x}(0),v_{y}(0)\}=\{0,0\}$,
  $e(0)=0$.}
\end{figure}
\begin{figure}[htbp]
\centerline{\includegraphics[width=7cm]{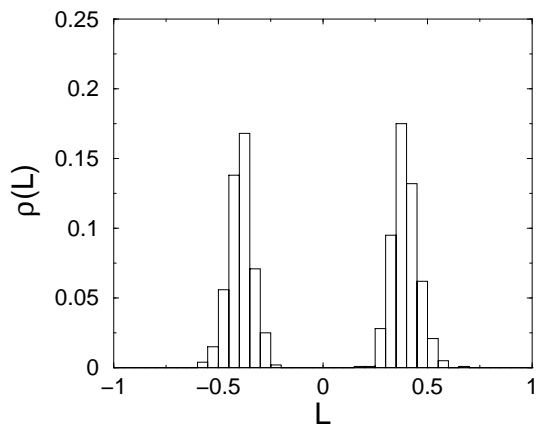}}
\caption[]{\label{distr_angular_momentum_N1000particles}
  Angular momentum distribution $\rho(L)$ of $N=1000$ Brownian agents
  after $t=150$.  The positive or negative sign of $L$ indicates the
  right- or lefthanded rotation. Parameters: $q_0=10.0$, $c=1.0$
  $\gamma=20.0$, $d_2=10.0$, $D=0.001$, $a=1.0$.}
\end{figure}

We now turn to the case of many, i.e. $i=1,...,N$ Brownian agents, which
is of importance for swarming. The dynamics of the multi-agent system is
then described by $2N$ coupled (stochastic) equations of the form
(\ref{model-2}). In this case, the computer simulations shown again
the characteristic rotational motion where, however, about half of the
agents rotate clockwise, while the other half rotates counterclockwise.
The two different cyclic directions can be clearly observed when looking
at the \emph{angular momentum} distribution, $\rho(L)$, where $\bbox{L}$
(for $m=1$) is defined as $\bbox{L}=\bbox{r}\times \bbox{v}$.  As
\pic{distr_angular_momentum_N1000particles} shows, this is a
\emph{bimodal} distribution of about equal height, indicating the both
left- and righthanded rotational directions with the same probability.
This is not surprising as long as independent random processes with a
certain symmetry are considered. But we use this graph here for
comparison with the results of the elaborated model, presented in the
next section. 

The simulation result also does not quite agree with the observation of
high density swarms of \textit{Daphnia}, which apparently cycle into
\emph{one}, i.e. the same direction. The reasons for this mismatch are
quite obvious: in our model, we have so far only considered
``point-like'' agents without any kind of mutual interaction, whereas in
real biological systems the coherent motion of the swarm is certainly
based on local interactions between the entities. The results of the
present model can be compared to the case of \textit{Daphnia} at low
density, where cycling around the light shaft in both directions is
observed~\citep{ordemann03:_pattern_formation} (compare
\pic{single_brownian_particle_wi_pump_wi_pot} with Fig. 3 of
~\citet{ordemann03:_motion_daphn_light_field} and
\pic{distr_angular_momentum_N1000particles} with Fig. 1a of
\citet{ordemann03:_pattern_formation}).  Thus, the question arises, which
kind of interaction may lead to the break in the rotational symmetry, as
observed in the \textit{Daphnia} experiments for high animal density.

\section{Modeling swarming with avoidance behavior}
\label{sec:avoidance-model}

\subsection{Interaction}
\label{sec:interaction}

So far, different forms of \emph{global} or \emph{local} interactions
have been introduced into swarming models. We mention

\begin{enumerate}
\item \emph{local} interactions via a self-consistent field that has been
  created by the agents and in turn influences their further movement
  and/or ``behavior'' \citep{fs-lao-family-97, helbing-fs-et-97,
    stevens-fs-97,couzin03:_self, erdmann05:_noise} -- chemotactic
  response is a prominent example here.
\item \emph{local} interactions based on the coupling of the agent's
  individual velocity to a local average velocity
  \citep{toner-tu-95,vicsek-et-95,levine-00,czirok-vicsek-00,couzin-02,chate-pre64,chate-prl92}.
\item \emph{global} interactions, such as the coupling of the agent's
  individual orientation (i.e.  direction of motion) to the mean
  orientation of the swarm \citep{czirok-et-96,czirok-vicsek-00}, or the
  coupling of the agent's individual position to the mean position
  (center of mass) of the
  swarm \citep{fs-eb-tilch-01,eb-fs-01-pas,mikhailov-zanette-99}
  , further couplings via the mean momentum or mean angular momentum or a
  combined set of invariants of motion
  \citep{czirok-et-96,fs-eb-tilch-01}.
\item interactions based on hydrodynamic coupling between agents
  \citep{erdmann03:_collec_motion_brown_partic_hydrod_inter}.
\end{enumerate}

Despite the fact that some of these models simulate coherent swarm
behavior or even rotation of the swarm in the same direction, there is
evidence that the underlying assumptions especially for global
interactions can hardly be satisfied by \emph{biological} observations,
thus their biological relevance is rather questionable. Therefore, in the
following section, we introduce local interactions between the agents
that indeed match with biological reality. In particular, we focus on a
special form of repulsive force between agents, which models avoidance
maneuvers between the agents. 

Experiments on \emph{Daphnia} swarming (\sect{sec:experimental-methods})
have shown that these animals tend to cycle into the same direction for
high \textit{Daphnia} densities. We argue that the reason for this may be
that animals try to avoid as much as possible collisions with other
animals -- which would occur much more frequently if different animals
cycled into opposite directions at the same time. Thus, a biologically
satisfactory assumption is to include \emph{avoidance behavior} in our
model of swarming, in order to test whether this would lead to the
observed break in the rotational symmetry described above.

\textit{Daphnia} are able to sense their environment to a certain degree
using their sensitive mechanoreceptors \citep{Gries:1999} and their
vision, i.e.  they can detect animals approaching them from the fore, and
then try to avoid collisions. In our models, we account for this by
assuming that there is a short-ranged repulsive force between agents, to
prevent their collisions. A similar idea was used by \citet{couzin03:_self} to 
describe collective motion in ants but, different from our approach, they
assume a hard-core repulsion in a fixed area around each agent. 

In our model, the repulsive force results from an \emph{interaction
  potential} $V(r_{i})$ around each agent $i$ that depends on its actual
position, $\bbox{r}_{i}$, and implicitly on both its actual velocity and
the velocity of the approaching agent.  In the following, two different
ansatzes for such an avoidance potential shall be introduced.  The first
one discussed in the next section, has the advantage of mathematical
simplicity, while the second one discussed in
\sect{sec:pedestr-avoid-model}, gives smoother trajectories of the
agents.  Both ansatzes, however, lead to the same dynamic behavior of the
swarm and therefore can be used equivalently.

\subsection{Simple Avoidance Model}
\label{sec:simple-avoid-model}

Our first approach is based on the assumption that the repulsion between
two agents depends inversely on the Euclidian distance
$r_{ij}=||\bbox{r}_{i j}||=||\bbox{r}_i-\bbox{r}_j||$ between two agents
$i$ and $j$ in two-dimensional space:
\begin{equation}
  \label{eq:simple_coulomb}
  V_{1}(r_{i j}) = \frac{c}{(r_{i j})^n} \ \ \ \ \ n\in
  \rez^+ \, ,
\end{equation}
where $c$ is some constant.  The force between two agents $i$ and $j$ can
be calculated as
\begin{equation}
  \label{eq:potential_force_relation}
  \bbox{f}_{i j}=-\bbox{\nabla} V_{1}(r_{i j})=\frac{c\,n}{(r_{i j }^{\epsilon})^{2 + n}}\,
  \bbox{r}_{i j} 
\end{equation}
Here, we have added a small offset $\epsilon$ to the denominator, to
avoid unwanted singularities if $r_{ij}\to 0$:
\begin{equation}
  \label{eq:r_alpha_beta_epsilon}
  r_{i j}^\epsilon = \sqrt{\epsilon + \bbox{r}_{i j} \cdot \bbox{r}_{i j}}
\end{equation}
In a $N$ agent system the total force on agent $i$ is simply the
sum over all 2 agent forces
\begin{equation}
  \label{eq:total_force}
  \bbox{F}_i=\sum_{j \neq i} \bbox{f}_{i j}
\end{equation}
The consideration of the avoidance behavior  leads to a modified equation
of motion, i.e.  \eqn{model-2} now reads
\begin{equation}
\frac{d}{dt}\bbox{r}_i =  \bbox{v}_i \;;\quad
\frac{d}{dt}\bbox{v}_i  =  - \gamma(v_{i}^{2})\,\bbox{v}_i
- a \bbox{r}_{i} + \sum_{i\neq j}\bbox{f}_{ij} + \sqrt{2D} \bbox{\xi}(t)
\label{langev-dep-extended}
\end{equation}
Again, we have assumed a \emph{linear} superposition of all these forces,
which seems to us the most simple assuption to start with. Other
assumptions for multiple interactions are of course possible, but hardly
motivated at the current stage. 

Simulations of \eqn{langev-dep-extended} with $\bbox{f}_{ij}$ defined by
\eqn{eq:potential_force_relation} show a swarming behavior with an
angular momentum distribution as in
\pic{distr_angular_momentum_N1000particles}. That means, we still find
left- and righthanded rotation at the same time and no symmetry break.
This is due to the fact that in \eqn{eq:simple_coulomb} the avoidance
behavior only depends on the distance between the two agents, i.e. the
repulsion is the same to the front and the rear for equal distances. This
assumption, however, can hardly be satisfied for \textit{Daphnia} because
they can mainly detect animals in front of them using their eye and the
mechanoreceptors at their swimming antennas. That means that the
direction of motion given by the velocity $\bbox{v}_i$ is crucial.

To account for this, we extend \eqn{eq:potential_force_relation} by
multiplying our force with an asymmetry factor $\omega_{i j}$ that
depends on the positions $\bbox{r}_i$, $\bbox{r}_j$ and the velocities
$\bbox{v}_i$, $\bbox{v}_j$ of the two agents $i$ and $j$ as explained
in the following.
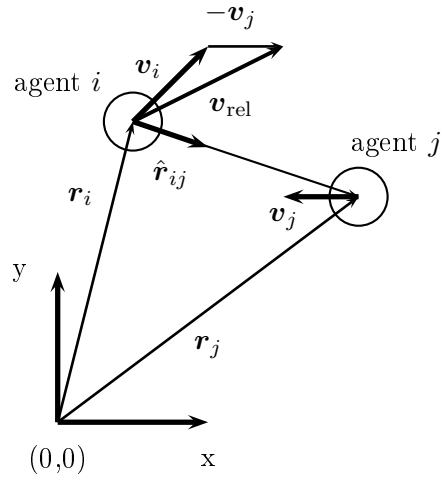
\begin{figure}[htbp]
\begin{center}
\makebox[0cm]{}\\[-0.3cm]
\begin{pspicture}(-1,-1)(6,6)
\pnode(0,0){or}
\pnode(2,0){xend}
\pnode(0,2){yend}
\ncline[linewidth=2pt]{->}{or}{xend}
\ncline[linewidth=2pt]{->}{or}{yend}
\rput(0,-0.5){(0,0)}
\rput(2,-0.5){x}
\rput(-0.5,2){y}
\pnode(1,4){ri}
\pnode(4,3){rj}
\pnode(2,5){vi}
\pnode(3,3){vj}
\pnode(3,5){vrel}
\pnode(2,3.66){eri}
\ncline[linewidth=1pt]{->}{or}{ri}
\rput(0.3,3.0){$\bbox{r}_i$}
\ncline[linewidth=1pt]{->}{or}{rj}
\rput(2.0,1.0){$\bbox{r}_j$}
\ncline{->}{ri}{rj}
\ncline[linewidth=2pt]{->}{ri}{vi}
\rput(1.2,4.7){$\bbox{v}_{i}$}
\ncline[linewidth=2pt]{->}{ri}{eri}
\rput(1.5,3.3){$\hat{\bbox{r}}_{ij}$}
\ncline[linewidth=2pt]{->}{rj}{vj}
\rput(3.0,2.7){$\bbox{v}_{j}$}
\ncline[linewidth=1pt]{->}{vi}{vrel}
\rput(2.3,5.4){$-\bbox{v}_{j}$}
\ncline[linewidth=1.5pt]{->}{ri}{vrel}
\rput(2.3,4.2){$\bbox{v}_{\rm rel}$}
\pscircle(1,4){0.4}
\pscircle(4,3){0.4}
\rput(4.5,3.7){agent $j$}
\rput(0,4.5){agent $i$}
\end{pspicture}
\end{center}
\caption[]{
  Visualization of the vectorial quantities associated with the agents.
  \label{vectors}}
\end{figure}

The prefactor $\omega_{ij}$ has to reflect two circumstances: (i) it must
increase with increasing relative velocity
\begin{equation}
  \label{eq:rel_velocity}
  \bbox{v}_{\rm rel} \equiv \bbox{v}_{i j}=\bbox{v}_{i}-\bbox{v}_j
\end{equation}
(see \pic{vectors}), because the agents would reach the point of
presumable collision faster and therefore the force to avoid this has to
be stronger; (ii) it has to consider whether two agents detect each other or
not. This is determined by the scalar product of the relative velocity, $
\bbox{v}_{\rm rel}$, and the unit vector
$\hat{\bbox{r}}_{ij}=\frac{\bbox{r}_{i j}}{||\bbox{r}_{i j}||}$ pointing
from agent $i$ towards agent $j$ (see \pic{vectors}).  Two agents $i$ and
$j$ detect each other only if
\begin{equation}
\label{eq:scalar_product_simple}
  \bbox{v}_{\rm rel} \cdot \hat{\bbox{r}}_{i j} > \eta \geq 0
\end{equation}
where $\eta$ accounts for the angle of perception, i.e. $\eta=0$ would mean a
visual angle of 180$\mathrm{^{o}}$ and $\eta>0$ corresponds to a smaller one.
To avoid singularities in $\hat{\bbox{r}}_{ij}$, we eventually replace
$||\bbox{r}_{i j}||$ by \eqn{eq:r_alpha_beta_epsilon}. Then, the prefactor
$\omega_{ij}$ reads in its final form:
\begin{equation}
  \label{eq:factor_scalarprod}
  \omega_{i j}^\epsilon = \left\{
    \begin{array}{ll}
\bbox{v}_{\rm rel} \cdot \hat{\bbox{r}}_{i j}^{\ \epsilon}&\ \ , \ \ \mbox{if}\ \ \ (\bbox{v}_{\rm rel}
\cdot \bbox{r}_{i j}) \geq \eta \ \ \ , \ \ \eta \geq 0 \\
\eta&\ \ , \ \ \mbox{else}
    \end{array}
 \right.
\end{equation}
where
\begin{equation}
  \label{eq:r_alpha_beta_epsilon_hat}
  \hat{\bbox{r}}_{i j}^{\ \epsilon}=\frac{\bbox{r}_{i j}}{r_{i j}^\epsilon}
\end{equation}
Considering the prefactor, the avoidance term $\bbox{f}_{ij}$,
\eqn{eq:potential_force_relation} reads now: 
\begin{equation}
  \label{eq:simple_scalar_prod_total_force}
  \bbox{f}_{i j}= \omega_{i j}^{\epsilon} \cdot \frac{c\,n}{(r_{i
  j}^\epsilon)^{2+n}} \bbox{r}_{i j}
\end{equation}
The equations of motion are still given by \eqn{langev-dep-extended}.
Computer simulations of this extended model
\eqs{langev-dep-extended}{eq:simple_scalar_prod_total_force},\ 
(\ref{eq:factor_scalarprod}) will be shown in \sect{sec:comp-simul-both}.

\subsection{Advanced avoidance model}
\label{sec:pedestr-avoid-model}

Our second approach is motivated by the observation that computer
simulations of the previous model, while showing the correct dynamic
behavior, have the visual disadvantage of abrupt turning maneuvers of the
agents. In order to improve the visual appearance, we adopt an ansatz for
the avoidance potential that has been originally used to model the
movement of pedestrians \citep{molnar95:_model_simul_dynam_fussg}:
\begin{equation}
  \label{eq:potential}
  V(R_{i}) = p\cdot \exp \left(- \frac{R_i(\bbox{r}_{ij},
\bbox{v}_{i},\bbox{v}_{j})}{\sigma} \right)
\end{equation}
$p$ denotes the strength and $\sigma$ the range of the potential, the
latter being a measure of the range of \emph{detection}.
$R_i$ is a specific function of the distance between agents, as explained
in the following.  Since all agents are moving, agent $i$ needs to
account for the space that will be occupied by all other agents $j$ in
the vicinity during the next time step. This space needed, depends both
on the agent's positions $\bbox{r}_{j}$ \emph{and} their velocity of
motion, $\bbox{v}_{j}$, so $R_i$ is a function of these. For further
specification, we introduce the unit vector in the direction of motion of
agent $i$, $ \bbox{n}_i^{0}=\bbox{v}_i/||\bbox{v}_i||$; $\bbox{n}_j^{0}$
is defined similarly. This allows to define a new
\emph{velocity-dependent} coordinate system for agent $i$, namely
$\bbox{y}_i$ and $\bbox{x}_i$ defined by:
\begin{equation}
  \label{eq:new_coordinate_system_y}
  \bbox{y}_i=\frac{v_i \bbox{n}_i ^{\ 0}-\delta  v_j
    \bbox{n}_j ^{\ 0}}{||v_i \bbox{n}_i ^{\ 0}-\delta v_j
\bbox{n}_j ^{\ 0} ||} \ \ \ \ ; \ \ \ \bbox{x}_i \perp \bbox{y}_i \ \ \
    \mbox{and} \ \ \ \left< \bbox{x}_i,\bbox{x}_i \right>=1 \ .
\end{equation}
If $\delta>0$, the direction of motion of agent $j$ is also taken into
account for agent $i$. The $\bbox{x}_i$ can be constructed by the
orthonomalization algorithm by \textsc{Gram-Schmidt}. Using this
coordinate system, the dependence of $R_i$ on the position and velocity
of agent $j$ is now given as
\begin{equation}
\label{eq:r}
R_i=\sqrt{\left< \bbox{r}_i-\bbox{r}_j,\bbox{x}_i \right>^2+\beta^2 \left< \bbox{r}_i-\bbox{r}_j,\bbox{y}_i \right>^2 }
\end{equation}
with a velocity-dependent function:
\begin{equation}
\label{eq:beta}
\beta = \left\{
  \begin{array}{ll}
\beta'&:\left< \bbox{r}_i-\bbox{r}_j,\bbox{y}_i \right> \geq 0\\
\displaystyle \frac{\beta'}{1+\lambda\cdot v_i}&:\left< \bbox{r}_i-\bbox{r}_j,\bbox{y}_i \right> < 0\\
  \end{array} \right.
\end{equation}
\begin{figure}[htbp]
\begin{center}
\centerline{
\includegraphics[width=3.5cm]{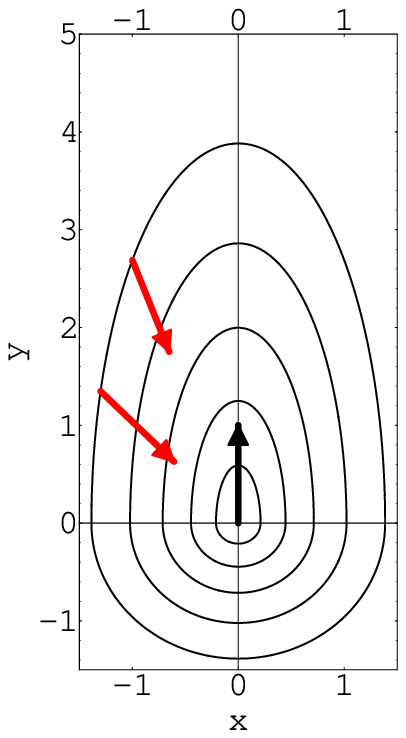}
\includegraphics[width=3.5cm]{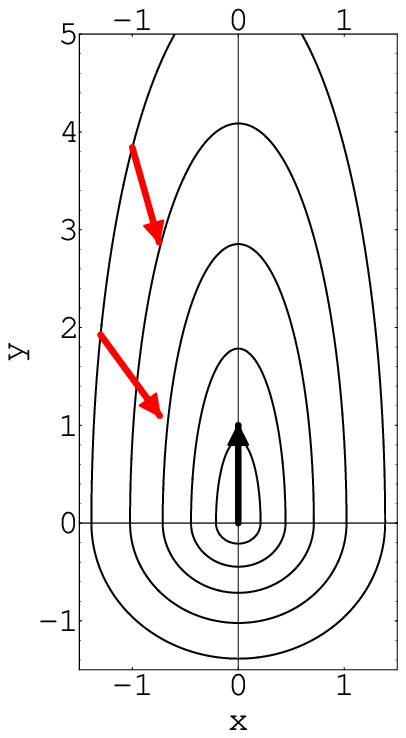}
\includegraphics[width=3.5cm]{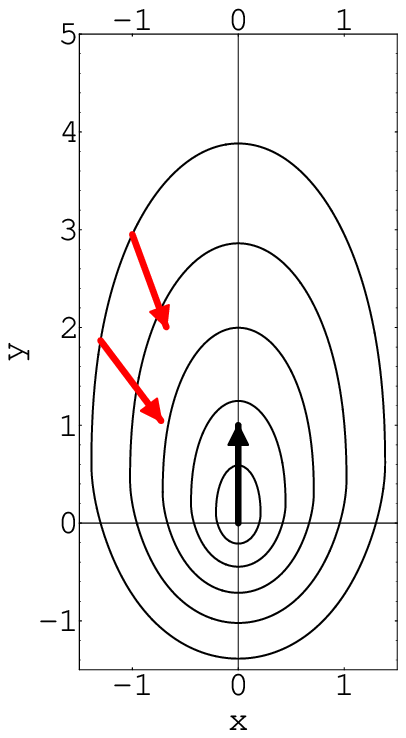}
}
\end{center}
\caption[]{\label{equi_potential_lines}
  Equipotential lines of the repulsive potential $V(r_{i})$,
  \eqn{eq:potential} for different parameters $\delta$ and $\lambda$.
  Left: $\delta=0$, $\lambda=1.8$, Middle: $\delta=0$, $\lambda=3$,
  Right: $\delta=0.4$, $\lambda=1.8$. The black arrow (in the center)
  indicates the agent in the origin, having a velocity of $v=\{0,1\}$.
  The two red (gray) arrows represent other agents with the same absolute
  value of the velocity (1) and point towards the origin. In the left,
  middle and right part of the figure, the equipotential lines shown
  correspond to the same values for the potential.
}
\end{figure}

In order to understand the meaning of the parameters $\beta'$ and
$\lambda$ we note that, if $\left< \bbox{r}_i-\bbox{r}_j,\bbox{y} \right>
\geq 0$ then agent $i$ is moving away from agent $j$.  This means that
increasing the value of $\beta'$ also means increasing $R_i$.  Therefore,
for both cases mentioned in \eqn{eq:beta}, $\beta'>1$ will lead to a
reduction in the repelling force between agents $i$ and $j$.  On the
other hand, an increase in $\lambda \cdot v_{i}$ means that the repelling
force between agents $i$ and $j$ will increase if agent $i$ is moving
towards $j$. We note that these assumptions lead to an asymmetric
repulsive potential $V(r_{i})$ around each agent.  The potential defined
by \eqn{eq:potential} with \eqn{eq:r} can be seen in
\pic{equi_potential_lines} for different parameters.

Eventually, with the known repulsive interaction potential $V(r_i)$,
the force between any two agents $i$ and $j$ is given as:
\begin{equation}
  \label{eq:force}
  \bbox{f}_{i j}=-\bbox{\nabla} V(R_{i}) = \frac{p}{\sigma \cdot R_i} \exp \left(- \frac{R_i}{\sigma}
  \right) (\bbox{r}_i-\bbox{r}_j)
\end{equation}
The dynamics of this model is also described by
\eqs{langev-dep-extended}{gamma-v2}.

\section{Results of computer simulations of both models}
\label{sec:comp-simul-both}

\subsection{Simulation of vortex formation}
\label{sec:sim-vortex}

The computer simulations of both, the \emph{simple avoidance model}
(\sect{sec:simple-avoid-model}) as well as the \emph{advanced avoidance
  model} (\sect{sec:pedestr-avoid-model}) show indeed the expected
symmetry break for the swarming behavior in agreement with the biological
observations. Spatial snapshots of a computer simulation of the
multi-agent system with respect to avoidance behavior, together with the
respective distribution of the angular momenta $\rho(L)$ are shown in
\pic{velocity_shots}. The results of both models can be concluded as
follows:
\begin{enumerate}
\item On the spatial level, we observe the emergence of a \emph{coherent
    motion} of the multi-agent swarm out of a random initial
  distribution.  This collective motion is characterized by a
  \emph{unique cycling direction} (either left- \emph{or} righthanded
  rotation).
\item We further observe the formation of a vortex which is rather
  similar to the \textit{Daphnia} swarm cycling round the light beam.
\item While in one simulation all agents cycle in the same direction, we
  note that in different simulations the cycling direction can be also
  opposite, i.e. there is \emph{no preferred cycling direction} for the
  swarm, which also agrees with the observations of the \textit{Daphnia}
  swarm.
\end{enumerate}

The computer simulations for the \emph{simple avoidance model}, while
showing the correct dynamic behavior, have the visual disadvantage of
abrupt turning maneuvers of the agents. To this end, the \emph{advanced
  avoidance model} was introduced in \sect{sec:pedestr-avoid-model}. From
the computer simulations of the \emph{advanced avoidance model} the
following improvements can be seen:
\begin{enumerate}
\item The movements of the agents look smoother, more \textit{Daphnia} like.
\item Although cycling in the same direction, there are some agents that
  come near the center.
\item For certain parameters a spontaneous change in the rotating
  direction can be observed. This occurs in particular if agent $i$ takes
  strongly into account ($\delta \approx 0.5$) the movement of agent $j$.
\end{enumerate}

\begin{figure}[htbp]
  {
\begin{minipage}{5cm}
\begin{center}
\centerline{\includegraphics[width=5cm]{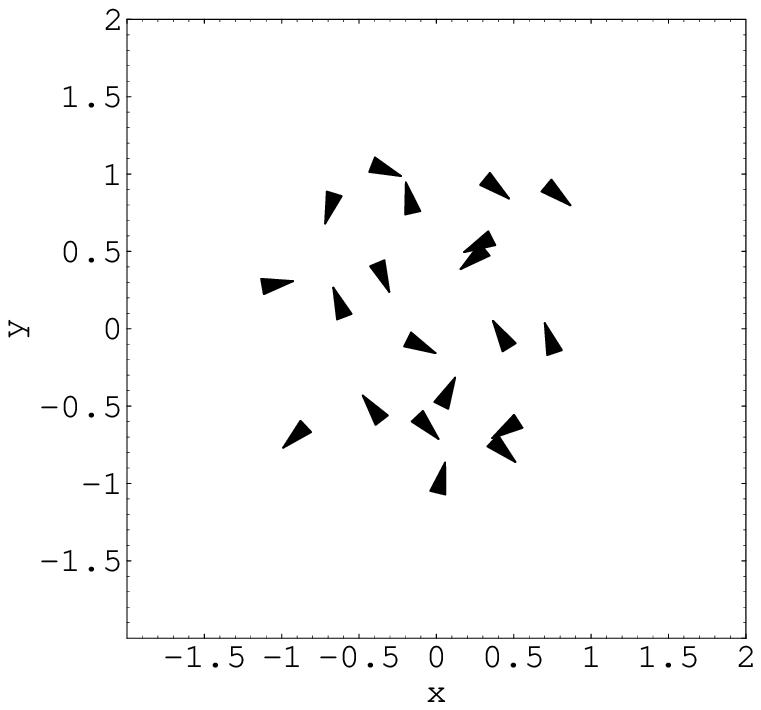}}
\end{center}
\end{minipage}
\hfill
\begin{minipage}{5cm}
\begin{center}
\centerline{\includegraphics[width=5cm]{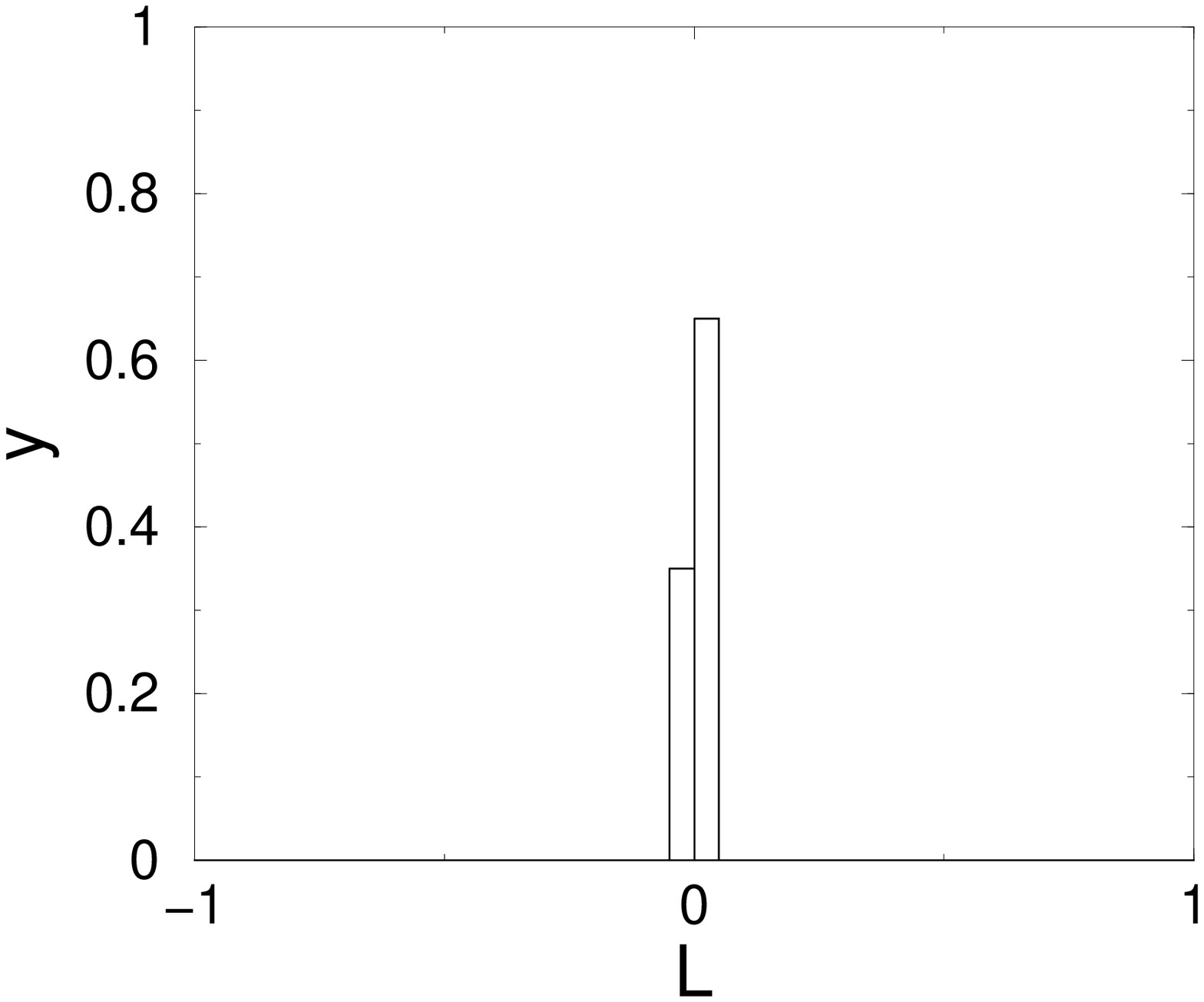}}
\end{center}
\end{minipage}
\vspace{-40pt}
{\begin{center} (a) \end{center}}
}
{
\begin{minipage}{5cm}
\begin{center}
\centerline{\includegraphics[width=5cm]{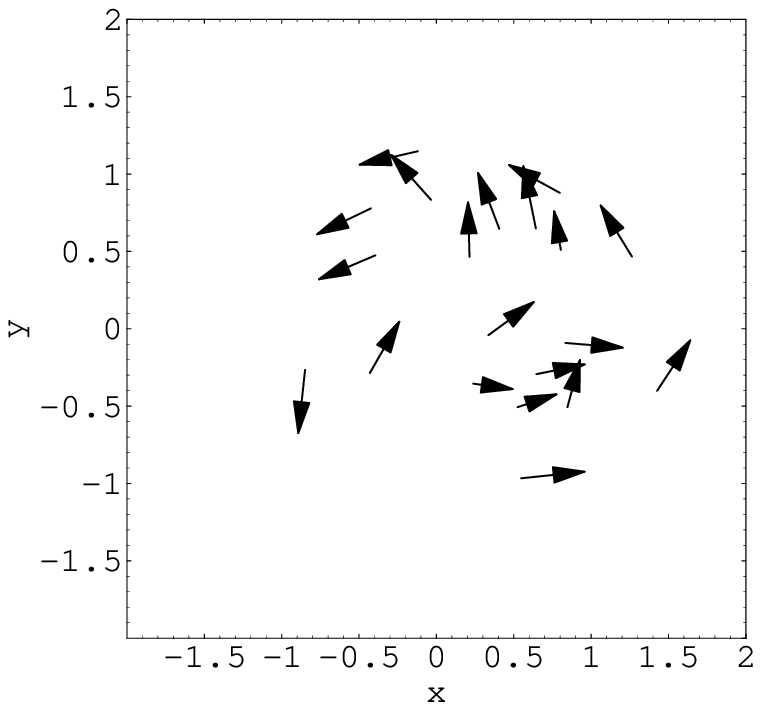}}
\end{center}
\end{minipage}
\hfill
\begin{minipage}{5cm}
\begin{center}
\centerline{\includegraphics[width=5cm]{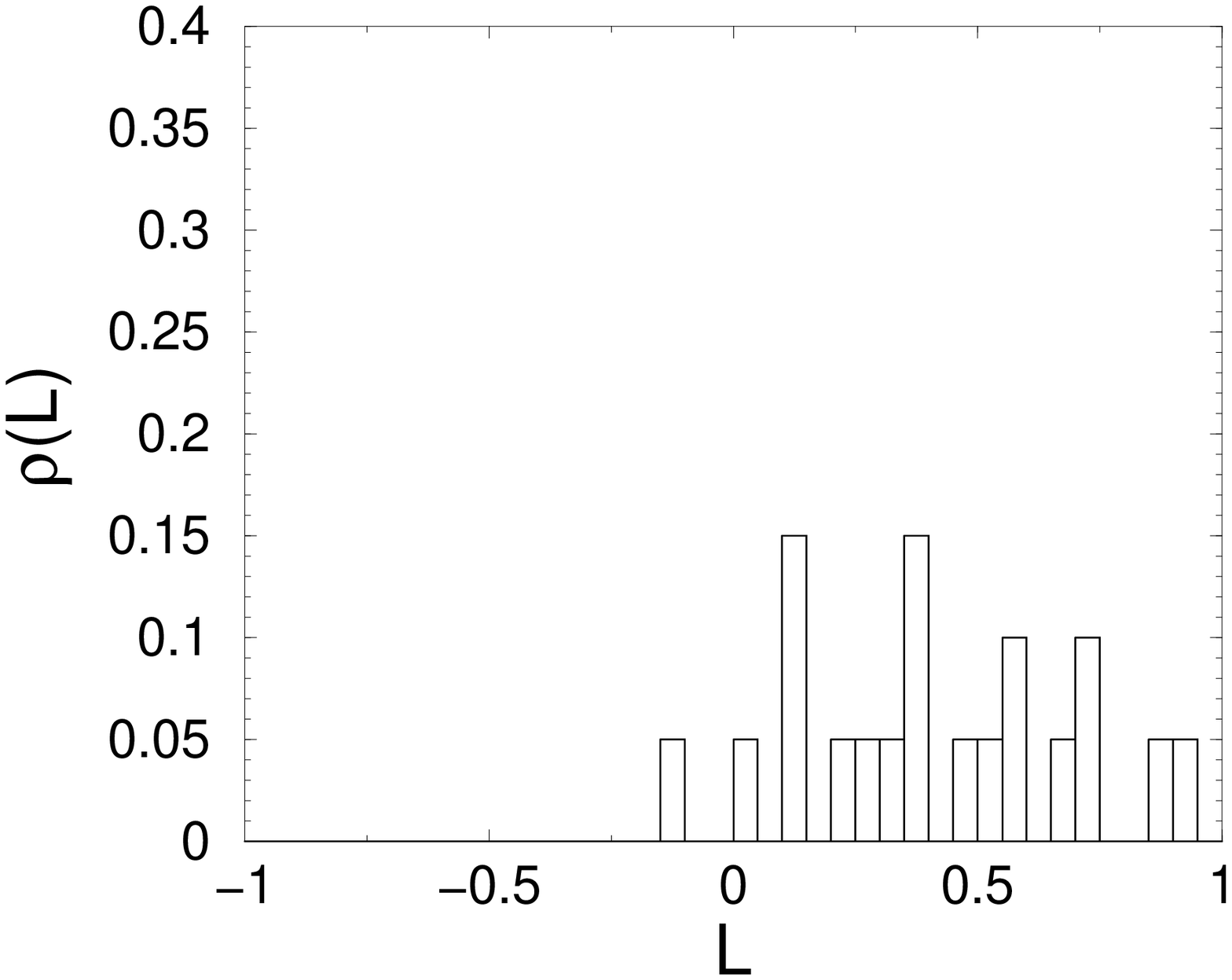}}
\end{center}
\end{minipage}
\vspace{-40pt}
{\begin{center} (b) \end{center}}
}
{
\begin{minipage}{5cm}
\begin{center}
\centerline{\includegraphics[width=5cm]{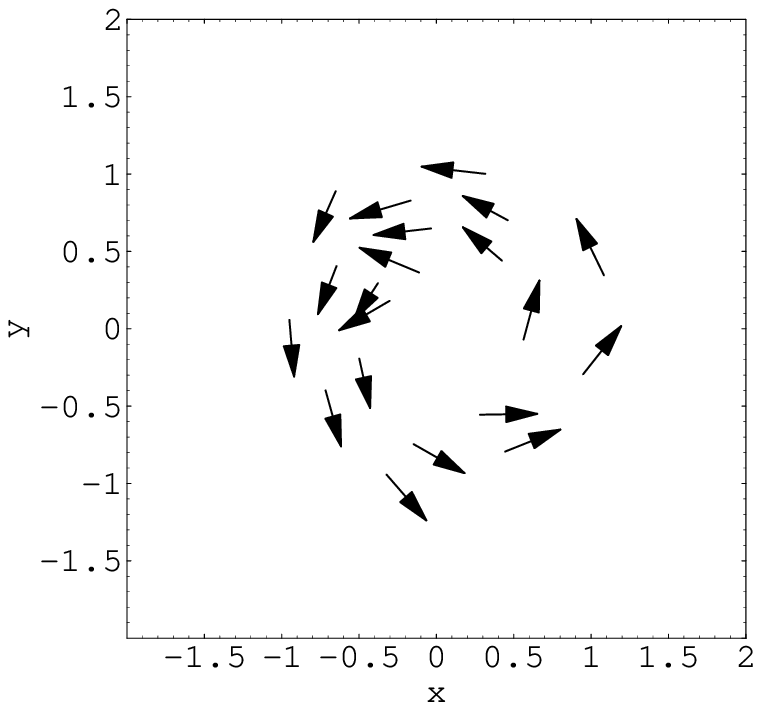}}
\end{center}
\end{minipage}
\hfill
\begin{minipage}{5cm}
\begin{center}
\centerline{\includegraphics[width=5cm]{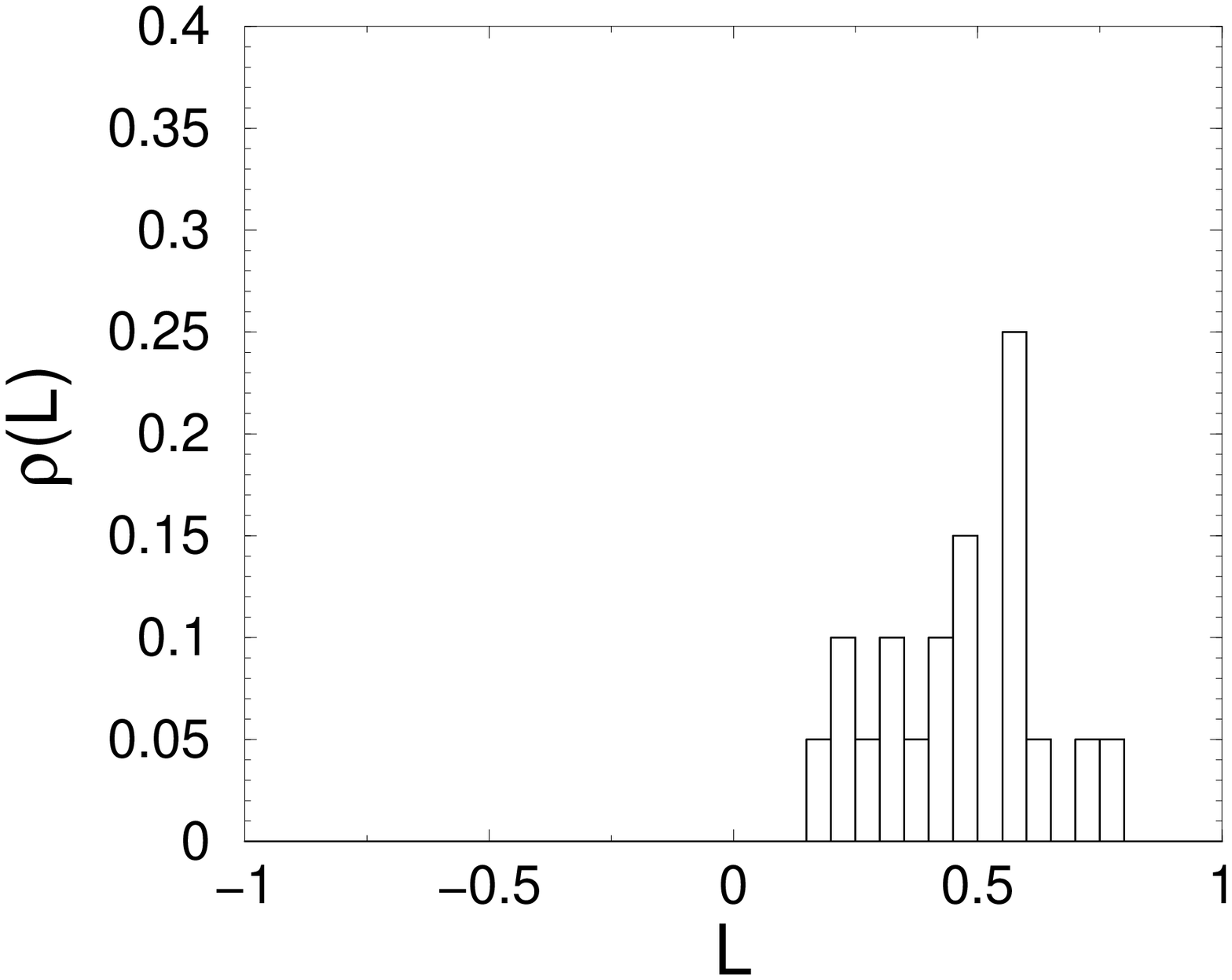}}
\end{center}
\end{minipage}
\vspace{-40pt}
{\begin{center} (c) \end{center}}
}
\caption{  \label{velocity_shots}
  Spatial snapshots (left) and distribution of angular momentum $\rho(L)$
  (right) for a multi-agent system ($N=20$) at three different times: (a)
  $t=0$ (b) $t=8$ and (c) $t=55$. The length of the arrows indicates the
  velocities.  Initial conditions: $\{x_i(0);y_i(0)\} \in [-1.5;+1.5]$, 
  $\{v_i^x(0),v_i^y(0)\}=\{0,0\}$, $e_i(0)=0$, parameters: $\gamma=20.0$,
  $d_2=20.0$, $q_0=10.0$, $c=1.0$, $D=0.005$, $a=1.0$ $p=0.8$,
  $\sigma=0.1$, $\delta=0.0$, $\lambda=10.0$, $\beta^{\prime}=2$.  A
  video of the computer simulations can be viewed at
  \texttt{http://intern.sg.ethz.ch/publications/2005/web-ms.html}.
}
\end{figure}

In order to demonstrate the influence of the avoidance interaction on the
collective motion of the swarm, we have conducted a computer simulation
where the interaction between agents is ``switched on'' at time $t=150$.
Thus, in the beginning, the swarm consists of non-interacting agents as
described in \sect{sec:agent-model-biol}.  In
\pic{fig:L_vs_t_w_ipot_on} the evolution of the angular momentum
distribution in time is shown.  In the very beginning, we find a broad
distribution of $\rho(L)$ centered around $L=0$.  This distribution
evolves towards a clear bimodal distribution as also shown in
\pic{distr_angular_momentum_N1000particles}, indicating the complete
symmetry between lefthanded and righthanded rotational direction. When
the interaction potential becomes effective at $t=150$, the agents start
to avoid collisions and thus tend to move into the same direction. This
can be clearly seen in \pic{fig:L_vs_t_w_ipot_on} where $\rho(L)$
transforms from a bimodal into an \emph{unimodal} distribution after
$t>150$. The transformation period ($\Delta t \approx 30$) is
characterized by large fluctuations that sometimes even give the less
frequent rotational direction a chance to take over.

\begin{figure}[htbp]
  \centerline{\includegraphics[width=8cm]{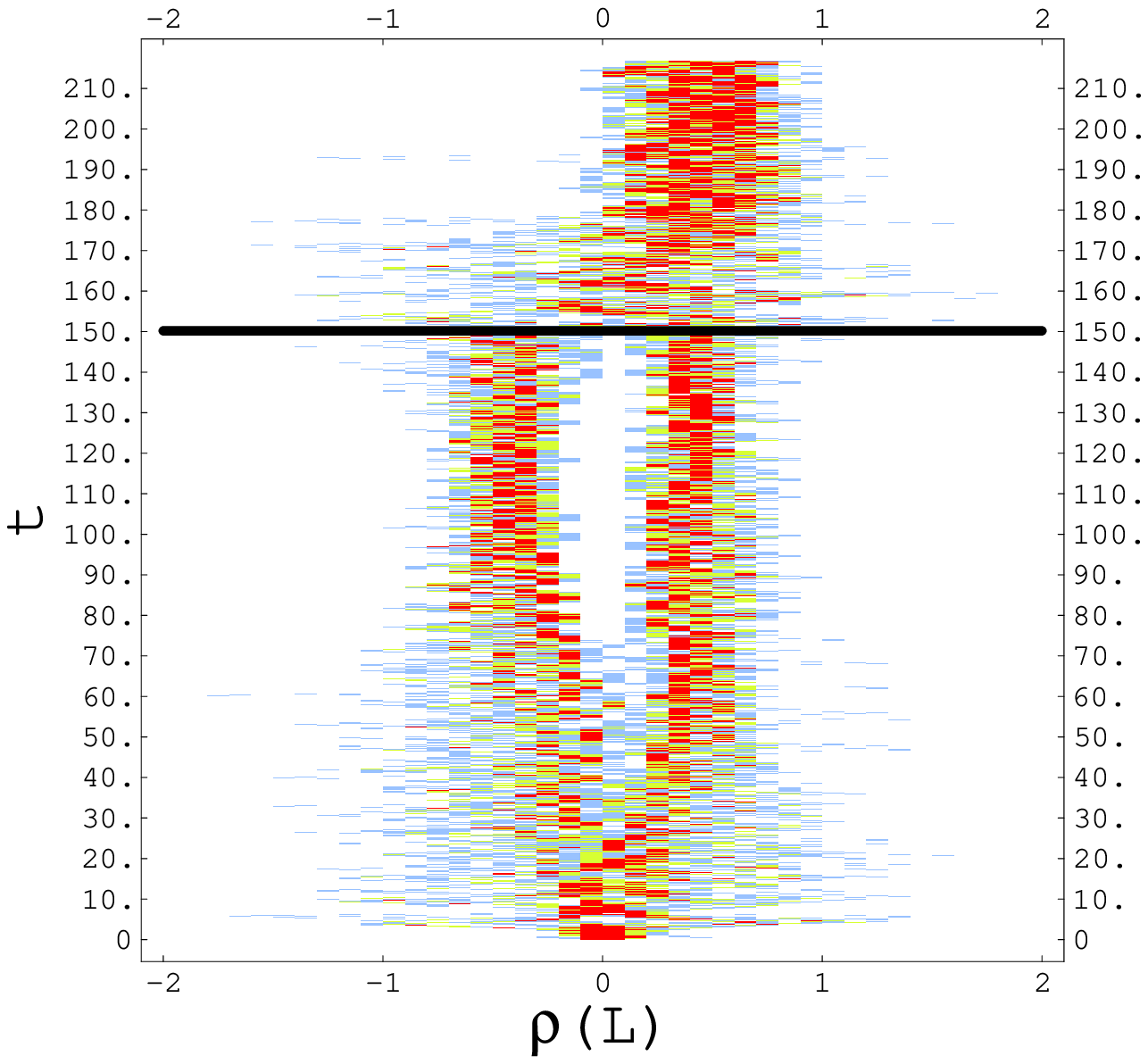}}
  \caption{Density plot (indicated by the gray scale)
    of the angular distribution $\rho(L)$ versus time.  The interaction
    between agents is turned on at $t=150$ (compare to
    \pic{distr_angular_momentum_N1000particles}). For the parameters and
    the setup see \pic{velocity_shots}.}
  \label{fig:L_vs_t_w_ipot_on}
\end{figure}

The symmetry break towards one cycling direction can be also interpreted
as a process of consensus formation in groups of animals as discussed by
\citet{couzin05:_effec}. They have developed a model to investigate the
effective leadership and decision making in animal groups in detail. To
this end, it was assumed that a small portion of the animals is
\emph{informed} and therefore has a desired direction of motion.  The
model is based on a mechanism with different tunable parameters, namely
the number of informed group members and the weight they give to their
preferred direction.  Through extensive computer simulations the authors
find that (a) for a given group size the accuracy of group motion
increases asymptotically as the proportion of the informed individuals
increases, and (b) the larger the group size, the smaller the proportion
of informed individuals needed.  \citet{couzin05:_effec} also investigate
how individuals achieve consensus about the majority direction in case of
different informed groups with competing directions.  A feedback
mechanism about adjusting the weights is proposed, to resolve this
conflict.

We like to point out that our model does not need such tunable model
parameters to achieve consensus about the moving direction. In fact, each
individual in our model, while moving, has only the desire to avoid
collisions with other animals. Thus, the collective motion emerges as a
self-organized phenomenon. To show that the transition from bimodal to
unimodal motion is inherent in our model, in the following section we
show how this transition occurs with increasing swarm size.

\subsection{Swarm size dependence of vortex formation}

So far, we have shown that our model can in fact reproduce the observed
cycling behavior of \emph{Daphnia} swarms. This means our proposed
avoidance behavior on the microscopic scale indeed leads to the symmetry
break on the macroscopic scale, whereas for single agents a symmetric
distribution of the angular momentum is found, in agreement with the
biological observation of \emph{Daphnia}.

This leads us to the question whether there is a critical swarm size at
which the \emph{emergence} of a vortex can be observed. In order to
investigate this, we have conducted extensive computer simulations of our
model with a fixed set of parameters, but different swarm sizes.  We
point out that the realization of a vortex swarm strongly depends on the
parameters of the model, in particular how much the symmetry break is
enforced by the avoidance potential. Thus, any conclusion about a
critical swarm size for vortex formation, drawn in the following, is
valid only for the particular parameter setting. This holds also for the
scaling relation discussed below.

Secondly, we note the strong dependence of the vortex formation on
stochastic influences. I.e., whether or not a vortex is formed, how long
it takes for the establishment of a common cycling direction, and what
this direction will be, is affected by stochastic fluctuations, which
play a considerable role especially for \emph{small} numbers of agents.

To compensate for this, we have measured the angular momentum
distribution of the swarm only after a sufficient time, $t=300$, where a
common cycling direction was established in all cases. This, however,
does not mean that \emph{all} agents follow the same direction at that
particular time. Further, there could still be large fluctuations
afterwards (as can be clearly observed in the computer simulation video
mentioned in \pic{velocity_shots}).  To account for this, we have
monitored the angular momentum distribution over the next 50 time units,
i.e.  between $t=300-350$, and have averaged over that time interval.  We
note that $t=300$ does not mean simulation steps, but physical time,
where the simulation interval was chosen as $\Delta t= 5\cdot 10^{-4}$,
i.e. $t=300$ corresponds to $6\cdot 10^{5}$ simulation steps and the
distribution was averaged over the next $10^{5}$ simulation steps.  It is
obvious that due to the pairwise interaction of the agents by means of
the avoidance potential, the computational effort for each simulation
step also increases with $N^{2}$.  Further, we have averaged the results
of computer simulations over 80 runs, from which we calculated the sample
standard deviation.

\pic{fig:phase_1} shows a function of the mean fraction of agents with a
particular angular momentum dependent on the swarm size $N$,
\begin{equation}
  \label{eq:frac}
\mathcal{F}_{L+}(N) = \abs{\bar{x}_{L+}-\bar{x}_{L-}}= 
\abs{2\bar{x}_{L+}-1}\;;\quad  \bar{x}_{L+}=\frac{1}{s\,m\,N} 
\sum_{k=1}^{s} \sum_{n=0}^{m} N^{(k)}_{L+}(t+n\Delta t)
\end{equation}
where $N_{L+}^{(k)}(t)$ is the number of agents found with a positive
angular momentum in simulation $(k)$ at time $t$. For the simulations
$s=10$, $t=300$, $\Delta t= 5\cdot 10^{-4}$ and $m=10^{5}$ were chosen.
\begin{figure}[htbp]
\bigskip
\centerline{
\includegraphics[width=9.0cm,angle=0]{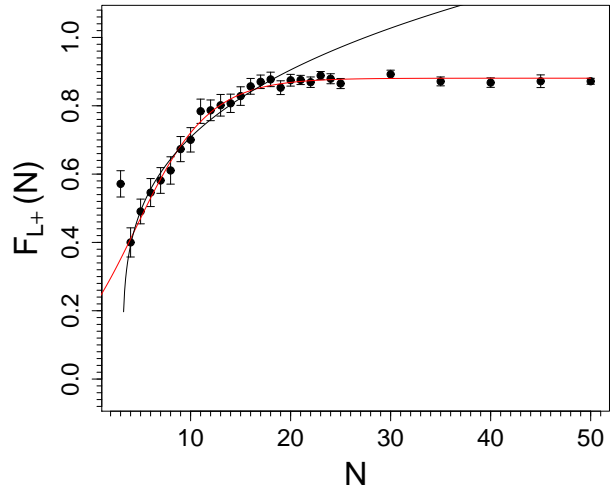}
}
\caption{\label{fig:phase_1} Mean fraction of agents with positive
  angular momentum $\mathcal{F}_{L+}(N)$, \eqn{eq:frac}, vs. swarm size
  $N$. The results are averaged over 80 simulations for each data point.
  The fitted curves are given by Eqs. \eqref{eq:vics}, \eqref{eq:fit}
  (black) and Eqs. \eqref{eq:fit2}, \eqref{eq:fit3} (red: saturation
  curve).  Parameters: $q_0=10.0$, $c=1.0$ $\gamma_0=20.0$, $d_2=10.0$,
  $D=0.001$, $a=1.0$, $p=0.1$, $\sigma=0.1$ , $\delta_\alpha=0.01$,
  $\lambda=10.0$.}
\end{figure}

If a swarm has a bimodal angular momentum distribution as shown in
Fig.  \ref{distr_angular_momentum_N1000particles}, $\bar{x}_{L+}$
would be 0.5 and $\mathcal{F}_{L+}(N)=0$, i.e. no \emph{common}
cycling direction has been established, that is followed by a majority
of agents. On the other hand, a clear unimodal distribution as e.g.
observed in Fig.  \ref{velocity_shots}, would lead to either
$\bar{x}_{L+}=0$ or 1, i.e.  $\mathcal{F}_{L+}(N)=1$. 

As Fig. \ref{fig:phase_1} indicates, the emergence of a common cycling
direction occurs between swarm sizes of 4 to 18 agents (for the given set
of parameters).  Even for large swarm sizes, the common cycling direction
is not followed by \emph{all} agents, a noticeable minority fraction
still cycles its own way. This, however, should be not considered as a
drawback of the model, in fact it makes it much more realistic, as the
computer simulation video also shows.  This behavior also agrees with
observations of \emph{Daphnia} swarms at high density. In particular,
animals at the border of the swarm still do not follow the cycling
direction of the majority.

We note that the occurence of a common cycling direction is not an abrupt
transition, but a gradual one. One can argue that this resembles a
kinetic phase transition analogous to the continuous phase transition in
equilibrium systems \citep{vicsek-et-95}, i.e.
\begin{equation}
  \label{eq:vics}
F_{L+}(N) = c_0 \, (N-N_{\rm c})^\kappa
\end{equation}
where $N_{\mathrm{c}}$ is the onset of the transition and $\kappa$ is the
critical exponent. For their model \citet{vicsek-et-95} have determined
$\kappa$ as 0.35. We have tested this finding against our simulation
results, as shown in Fig. \ref{fig:phase_1} and found for the least
square fit and standard errors in the region $N\in [4,20]$:
\begin{equation}
  \label{eq:fit}
  N_c=3.240\pm 0.328; \; c_0 =0.422\pm 0.029; \;
  \kappa=0.270\pm 0.025
\end{equation}
On the first glimpse, there is a close similarity between
\citet{vicsek-et-95} and our findings with respect to the value of the
critical exponent $\kappa$. I.e., we emphasize that vortex formation
under repulsion also occurs at a critical point.

However, as Fig. \ref{fig:phase_1} clearly shows, the scaling suggested
by \citet{vicsek-et-95} only holds for a restricted range of $N$ - which
by the way was already the case in the original paper (cf. Fig 2b of
\citet{vicsek-et-95}). A much better fit of the observed transition is
provided by the following function:
\begin{equation}
  \label{eq:fit2}
  F_{L+}(N) =\kappa \left[
    1 + \exp{ \left(-\frac{N -
          N_c}{\tau} \right)}
  \right]^{-1} 
\end{equation}
with the best-fit parameters and standard errors: 
\begin{equation}
  \label{eq:fit3}
  N_c=4.496\pm 0.136;\; \tau=3.651\pm 0.161;\; 
\kappa=0.880\pm 0.004 
\end{equation}
As shown in Fig. \ref{fig:phase_1}, this function fits the simulated data
extremely well. The only outlier for $N=3$ is due to the fact, that for 3
agents at least 2 agents go in the same direction. That means that the
probability of breaking the symmetry is a lot higher in this particular
case.

In addition to the depencence on the swarm size or density,
\citet{vicsek-et-95} also discussed a phase transition with respect to
noise, $\eta$, in the \emph{alignment} of the individuals and suggested a
scaling similar to Eqn. (\ref{eq:vics}). As the critical value,
$\eta_{\mathrm{c}}$, depends on the lattice size of their computer
simulations, there is no direct comparison with this result possible.
However, we note the obvious similarities to other types of structure
formation which also occur only below a certain critical temperature.
This holds also for our model where a common cycling direction can emerge
only below a critical noise level. As a difference to the investigations
of \citet{vicsek-et-95}, noise in our model enters the equation of
motion, Eqn. \eqref{langev-dep-extended}, whereas the avoidance
interaction is without noise.

\section{Model testing of avoidance maneuvers}
\label{sec:gath-empir-evid}

So far, reasonable assumptions about local interactions of \emph{Daphnia}
(such as local repulsion) have been taken into account. But it still
remains to be tested empirically and theoretically whether avoidance
maneuvers of \textit{Daphnia} are really responsible for the symmetry
break observed in \textit{high density} swarms.

To gather evidence on the microscopic dynamics of our model, we used data
provided by Anke Ordemann \citep{ordemann03:_unpublished} about avoidance
maneuvers of pairs of \textit{Daphnia} approaching each other in a
horizontal plane. Snapshots of a typical avoidance time series are shown
in \pic{fig:avoidance_track}. A thorough analysis of similar experiments
would allow to determine the parameters in our model, but the time series
presented was the only data made available to us. Nevertheless, this
sequence already enables us to qualitatively compare the experimental
findings with our theoretical model of avoidance behavior.

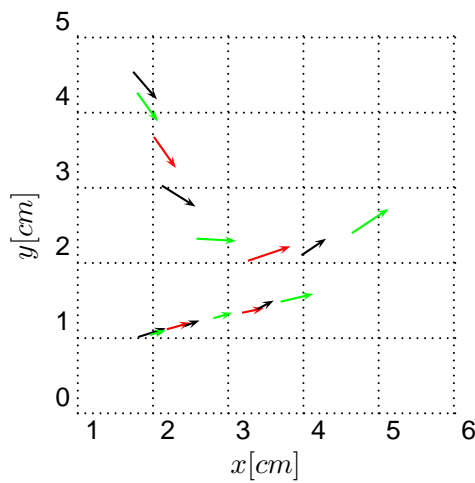
\begin{figure}[htbp]
\begin{center}
{\psset{unit=1cm}
\begin{pspicture}(1,-0.5)(6,5)
\psgrid[subgriddiv=1,griddots=10](1,0)(1,0)(6,5)
\rput(3.5,-0.7){$x [cm]$}
\rput{90}(0.3,2.5){$y [cm]$}

\psline[linecolor=black]{<-}(2.05263,4.17647)(1.73684,4.54412)
\psline[linecolor=green]{<-}(2.06579,3.88235)(1.78947,4.26471)
\psline[linecolor=red]{<-}(2.30263,3.26471)(2.01316,3.67647)
\psline[linecolor=black]{<-}(2.56579,2.75)(2.11842,3.02941)
\psline[linecolor=green]{<-}(3.10526,2.29412)(2.57895,2.32353)
\psline[linecolor=red]{<-}(3.82895,2.22059)(3.26316,2.02941)
\psline[linecolor=black]{<-}(4.30263,2.32353)(3.97368,2.10294)
\psline[linecolor=green]{<-}(5.13158,2.72059)(4.64474,2.39706)

\psline[linecolor=black]{<-}(2.17105,1.13235)(1.80263,1.01471)
\psline[linecolor=green]{<-}(2.17105,1.10294)(1.96053,1.04412)
\psline[linecolor=red]{<-}(2.5,1.20588)(2.18421,1.11765)
\psline[linecolor=black]{<-}(2.61842,1.23529)(2.42105,1.16176)
\psline[linecolor=green]{<-}(3.05263,1.33824)(2.80263,1.26471)
\psline[linecolor=red]{<-}(3.47368,1.39706)(3.18421,1.33824)
\psline[linecolor=black]{<-}(3.60526,1.5)(3.38158,1.38235)
\psline[linecolor=green]{<-}(4.13158,1.58824)(3.69737,1.48529)

\end{pspicture}
}
\end{center}
\caption[]{Experimental observation of an inter-animal avoidance maneuver
  of two \textit{Daphnia} in the horizontal plane
  ~\citep{ordemann03:_unpublished}. In our representation of the
  available data, each arrow represents the spatial orientation (head and
  tail) of the \textit{Daphnia} at successive times (every 0.2 sec.). In
  the sequence, same arrow colors (gray scales) correspond to the same
  time.  \label{fig:avoidance_track} }
\end{figure}

\begin{figure}[htbp]
\begin{center}
\hfill
\scalebox{1.1}{
{\psset{unit=1.5cm}
\begin{pspicture}(-2.4,-2.4)(1.5,1.5)
\psgrid[subgriddiv=1,griddots=10](1,1)(-2,-2)(1,1)
\rput(-0.5,-2.2){$x$}
\rput{90}(-2.2,-0.5){$y$}

\psline[linecolor=black]{->}(-1.99461,-0.99720)(-1.90987,-0.94411)
\psline[linecolor=green]{->}(-1.98498,-0.99395)(-1.88821,-0.96876)
\psline[linecolor=red]{->}(-1.91775,-0.98084)(-1.81933,-0.96316)
\psline[linecolor=black]{->}(-1.74104,-0.94866)(-1.64286,-0.92969)
\psline[linecolor=green]{->}(-1.64727,-0.93206)(-1.54828,-0.91791)
\psline[linecolor=red]{->}(-1.51035,-0.91060)(-1.41231,-0.89090)
\psline[linecolor=black]{->}(-1.38343,-0.88472)(-1.28548,-0.86457)
\psline[linecolor=green]{->}(-1.26453,-0.86201)(-1.16613,-0.84417)
\psline[linecolor=red]{->}(-1.13221,-0.84410)(-1.03289,-0.83245)
\psline[linecolor=black]{->}(-1.00911,-0.82660)(-0.91034,-0.81096)
\psline[linecolor=green]{->}(-0.88403,-0.81034)(-0.78441,-0.80158)
\psline[linecolor=red]{->}(-0.75451,-0.79928)(-0.65482,-0.79138)
\psline[linecolor=black]{->}(-0.63007,-0.78753)(-0.53047,-0.77863)
\psline[linecolor=green]{->}(-0.50112,-0.77769)(-0.40151,-0.76886)
\psline[linecolor=red]{->}(-0.37578,-0.76623)(-0.27638,-0.75533)
\psline[linecolor=black]{->}(-0.24973,-0.75259)(-0.14994,-0.74611)
\psline[linecolor=green]{->}(-0.12385,-0.74531)(-0.02392,-0.74167)
\psline[linecolor=red]{->}(0.00465,-0.74437)(0.10464,-0.74584)
\psline[linecolor=black]{->}(0.13046,-0.74602)(0.23045,-0.74461)
\psline[linecolor=green]{->}(0.25499,-0.74155)(0.35460,-0.73270)

\psline[linecolor=black]{->}(-1.97214,-0.06865)(-1.93051,-0.15958)
\psline[linecolor=green]{->}(-1.95611,-0.09923)(-1.89131,-0.17540)
\psline[linecolor=red]{->}(-1.89034,-0.16139)(-1.81706,-0.22944)
\psline[linecolor=black]{->}(-1.77116,-0.27014)(-1.69882,-0.33918)
\psline[linecolor=green]{->}(-1.69953,-0.33410)(-1.62553,-0.40137)
\psline[linecolor=red]{->}(-1.59296,-0.42285)(-1.51271,-0.48251)
\psline[linecolor=black]{->}(-1.49427,-0.49027)(-1.40952,-0.54335)
\psline[linecolor=green]{->}(-1.38328,-0.55161)(-1.29343,-0.59551)
\psline[linecolor=red]{->}(-1.26729,-0.59778)(-1.17283,-0.63060)
\psline[linecolor=black]{->}(-1.14800,-0.62718)(-1.04868,-0.63882)
\psline[linecolor=green]{->}(-1.02114,-0.63393)(-0.92120,-0.63034)
\psline[linecolor=red]{->}(-0.89818,-0.61947)(-0.79983,-0.60138)
\psline[linecolor=black]{->}(-0.77355,-0.59098)(-0.67751,-0.56313)
\psline[linecolor=green]{->}(-0.65683,-0.54977)(-0.56382,-0.51305)
\psline[linecolor=red]{->}(-0.53710,-0.50146)(-0.44598,-0.46025)
\psline[linecolor=black]{->}(-0.42483,-0.44688)(-0.33580,-0.40133)
\psline[linecolor=green]{->}(-0.31417,-0.38581)(-0.22837,-0.33445)
\psline[linecolor=red]{->}(-0.20613,-0.31636)(-0.12418,-0.25906)
\psline[linecolor=black]{->}(-0.10506,-0.24256)(-0.02820,-0.17859)
\psline[linecolor=green]{->}(-0.00692,-0.16270)(0.06986,-0.09863)
\end{pspicture}
}
}
\hfill
\scalebox{1.1}{
{\psset{unit=1.5cm}
\begin{pspicture}(-2.4,-2.4)(1.5,1.5)
\psgrid[subgriddiv=1,griddots=10](1,1)(-2,-2)(1,1)
\rput(-0.5,-2.2){$x$}
\rput{90}(-2.2,-0.5){$y$}

\psline[linecolor=black]{->}(-2.00000,-1.00000)(-1.91056,-0.95528)
\psline[linecolor=green]{->}(-1.99189,-0.99750)(-1.89290,-0.98331)
\psline[linecolor=red]{->}(-1.98313,-0.99274)(-1.89064,-0.95473)
\psline[linecolor=black]{->}(-1.86061,-0.93478)(-1.76997,-0.89253)
\psline[linecolor=green]{->}(-1.73563,-0.87809)(-1.64334,-0.83959)
\psline[linecolor=red]{->}(-1.64590,-0.84393)(-1.55263,-0.80786)
\psline[linecolor=black]{->}(-1.50694,-0.79159)(-1.41331,-0.75647)
\psline[linecolor=green]{->}(-1.39569,-0.74836)(-1.30325,-0.71022)
\psline[linecolor=red]{->}(-1.27960,-0.69869)(-1.18866,-0.65710)
\psline[linecolor=black]{->}(-1.16156,-0.64777)(-1.06915,-0.60955)
\psline[linecolor=green]{->}(-1.04510,-0.59585)(-0.95459,-0.55332)
\psline[linecolor=red]{->}(-0.93031,-0.53955)(-0.84040,-0.49578)
\psline[linecolor=black]{->}(-0.81746,-0.48386)(-0.72752,-0.44014)
\psline[linecolor=green]{->}(-0.70188,-0.42795)(-0.61221,-0.38370)
\psline[linecolor=red]{->}(-0.59137,-0.36973)(-0.50306,-0.32282)
\psline[linecolor=black]{->}(-0.47665,-0.31142)(-0.38602,-0.26916)
\psline[linecolor=green]{->}(-0.36310,-0.25959)(-0.27223,-0.21783)
\psline[linecolor=red]{->}(-0.30237,-1.06039)(-0.20549,-1.08519)
\psline[linecolor=black]{->}(-0.18166,-1.08881)(-0.08385,-1.10966)
\psline[linecolor=green]{->}(-0.05467,-1.11362)(0.04438,-1.12742)
\psline[linecolor=red]{->}(0.06929,-1.12665)(0.16885,-1.13599)

\psline[linecolor=black]{->}(-2.00000,0.00000)(-1.96286,-0.09285)
\psline[linecolor=green]{->}(-1.96166,-0.08766)(-1.91243,-0.17470)
\psline[linecolor=red]{->}(-1.93877,-0.12091)(-1.87790,-0.20025)
\psline[linecolor=black]{->}(-1.83480,-0.24071)(-1.76732,-0.31451)
\psline[linecolor=green]{->}(-1.75191,-0.32663)(-1.67998,-0.39610)
\psline[linecolor=red]{->}(-1.66432,-0.40272)(-1.58685,-0.46596)
\psline[linecolor=black]{->}(-1.55892,-0.48705)(-1.47901,-0.54717)
\psline[linecolor=green]{->}(-1.45734,-0.55852)(-1.37298,-0.61222)
\psline[linecolor=red]{->}(-1.35418,-0.62504)(-1.26919,-0.67774)
\psline[linecolor=black]{->}(-1.24347,-0.69050)(-1.15628,-0.73948)
\psline[linecolor=green]{->}(-1.12929,-0.74819)(-1.03855,-0.79021)
\psline[linecolor=red]{->}(-1.01695,-0.80327)(-0.92705,-0.84706)
\psline[linecolor=black]{->}(-0.90218,-0.85610)(-0.81161,-0.89849)
\psline[linecolor=green]{->}(-0.78537,-0.90471)(-0.69184,-0.94010)
\psline[linecolor=red]{->}(-0.66646,-0.94891)(-0.57216,-0.98219)
\psline[linecolor=black]{->}(-0.54388,-0.99025)(-0.44878,-1.02115)
\psline[linecolor=green]{->}(-0.30715,-0.23450)(-0.21665,-0.19195)
\psline[linecolor=red]{->}(-0.19006,-0.17765)(-0.10007,-0.13403)
\psline[linecolor=black]{->}(-0.07864,-0.12647)(0.01289,-0.08621)
\psline[linecolor=green]{->}(0.03814,-0.07595)(0.13029,-0.03710)
\psline[linecolor=red]{->}(0.15384,-0.03012)(0.24641,0.00770)

\end{pspicture}
}
}
\hfill
\end{center}
 \caption[]{
   Left: Simulation of an inter-agent avoidance maneuver in the
   horizontal plane.  Each arrow represents the spatial orientation
   (drawn from the velocity vector) of the agent at successive
   times (every 0.2 time units). In the sequence, same arrow colors (gray
   scales) correspond to the same time. Parameters: $q_0=10.0$, $c=1.0$
   $\gamma_0=20.0$, $d_2=10.0$, $D=0.001$, $p=0.2$, $\sigma=0.2$ ,
   $\delta_\alpha=0.1$, $\lambda=10.0$, $a=0.1$, Right: Simulation with
   the same setting but without avoidance interaction, $p=0.0$.
   \label{fig:avoidance_sim}}
\end{figure}
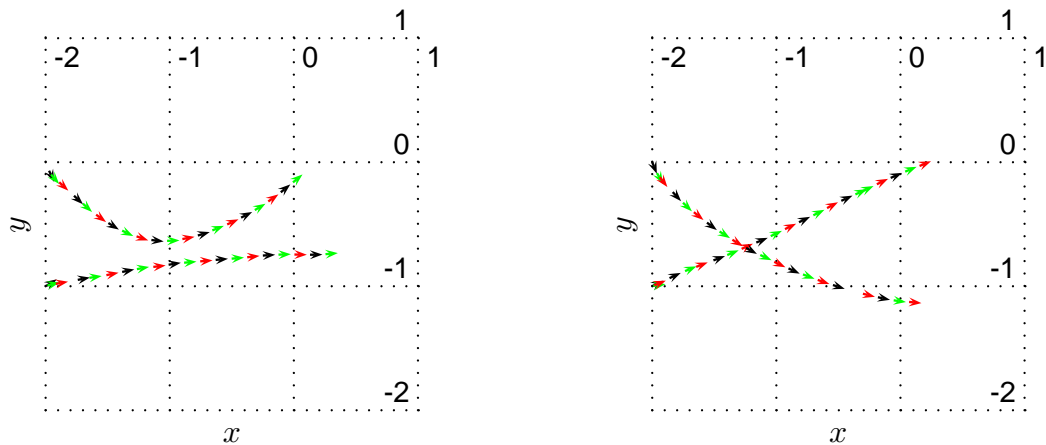

To this end, we have simulated the avoidance maneuver of two agents in a
situation similar to the one shown in Fig.  \ref{fig:avoidance_track}.
From the simulation results shown in the left part of Fig.
\ref{fig:avoidance_sim} it can be clearly seen that the two agents avoid
each other not in a symmetrical way, but rather similar to the
experimental findings.  One may argue that this is due to the additional
influence of the environmental potential which may more affect the agent
closer to the origin $(0,0)$. In order to test this, we have simulated
the same situation without the avoidance interaction potential, but just
the environmental potential. The result, shown in the right part of Fig.
\ref{fig:avoidance_sim}, clearly indicates that the environmental
potential has only very little affects the agent's trajectories and thus
cannot be responsible for the realistic avoidance behavior shown in the
left part of Fig.  \ref{fig:avoidance_sim}.

Comparing the simulation of the model with the above experimental
observations in {\it Daphnia}, it becomes obvious that inter-animal
avoidance maneuvers similar to the ones incorporated in the current
Brownian agent model are most likely involved in the mechanism that
causes symmetry breaking for high \textit{Daphnia} density and leads to
the observed vortex swarms.

\section{Discussion}
\label{disc}

In this paper, we tried to understand the vortex swarming behavior
observed in \textit{Daphnia} from rather \emph{minimalistic} assumptions
that, however, should have a clear \emph{biological relevance}. To this
end, we introduced a multi-agent model based on the concept of Brownian
agents. Different from other modeling approaches which are based e.g. on
cellular automata, our model considers both \emph{continuous} time and
space. Further, in addition to external and interaction forces effecting
the agent's behavior, we consider stochastic influences resulting e.g.
from random events or fluctuations in the environment. As a difference to
multi-agent models developed in artificial life, our model further allows
for a more specific discussion of the different parameters affecting the
agent's motion. It can be also seen as an alternative to some of the
self-propelled particle models \citep{chate-prl92, chate-pre64,
  vicsek-et-95} where alignment between individuals is explicit.  In our
model, we demonstrate that vortex swarming does not require an explicit
alignment rule for individuals but can be achieved through a combination
of attraction and repulsion. This insight allows us to reduce assumptions
about the cognitive requirements of swarming organisms.

At the end, we give a critical discussion of the Brownian agent model in
the biological context of \emph{Daphnia} swarming, in order to point out
its limitations and to give some hints for experimental verifications. 

The equation of motion for the Brownian agents is a generalized Langevin
\eqn{model-2}, which models the motion of the animals as continuous in
space and time.  The movement of single \emph{Daphnia}, however, consists
of a series of jumps as can be also observed in experimental
observations. This can at least be partially covered by using a larger
time step in our simulation, resulting in larger jumps during each time
interval. We argue that this would not change the overall dynamic
behavior of our model.  Thus, we kept the continuous approximation of the
discrete jumps and the small time step to avoid numerical instabilities.

Our theoretical description is based on the assumption of a
quasistationary energy depot of the agents.
This implies a constant take-up of energy, $q_{0}$, whereas
\emph{Daphina} feed in cycles while they do not move actively.  Thus, a
switch-like change in behavior seems to be more appropriate. This can be
covered in our model by assuming the take-up rate of energy a
time-dependent function, $q(t)$, and replacing the continuous
acceleration term in \eqn{model-2}, $d_{2}e_{i}(t)$, by a more complex
term that reflects a switch-like change between starvation and active
motion.
  
Response to external forces and force-based interaction between the
agents are two basic ingredients of our model of Brownian agents.
Interaction of biological entities, on the other hand, is often driven by
internal and behavioral reasons. This certainly sets the limits for our
approach which is inspired by, and based of, physical considerations
using minimal assumptions. Physics, per se, has no concept of
``behavior'' based on internalized motivations of an agent.  Thus, in our
model every kind of internal ``driving forces'' has to be
\emph{externalized} by assuming that the agent behaves as it would follow
an external force that leads it to the desired behavior. Such an approach
does not claim that these forces really exist in the outer
world, 
it is rather a convenient modeling formalism that allows to apply the
concepts of physics to the much more complex behavior of animals. An
alternative could be concepts from artificial intelligents, where the
\emph{internal} dynamics of agents is modeled explicitely (the so-called
BDI (belief-desire-intention) agents, for example, can have their own
internal world view).

In our model of Brownian agents, we could demonstrate the emergence of a
vortex swarm based on local asymmetric interactions of the agents.  The
global dynamic behavior was also found for other models (see also Sect.
\ref{sec:interaction}) exploiting other mechanisms of symmetry breaking.
We do not want to argue here about the most simplest one - at the end,
one has to find a compromise between ``simple'' and biologically
``satisfied''. Many of the proposed mechanisms -- such as the
hydrodynamic coupling -- still lack an experimental justification.  Our
avoidance model, on the other hand, could be at least visually tested by
comparing the experimental observations of two \emph{Daphnia} approaching
each other with a similar situation from the computer simulations.

Noteworthy, in our model the emergence of a vortex swarm is \emph{not}
enforced by the alignment of the agents, as used in other models.
Instead, we have included only the simple assumption of collision
avoidance. The specific form of our avoidance potential penalizes mainly
the head-head collision and thus promotes the dispersion of the agents
(and not the alignment).  As a consequence, to obtain a vortex swarm, an
additional attraction force is needed, which in the considered case
results from the \emph{environmental potential} (attraction towards the
center).  This is not a drawback of the model, but justified by the real
experimental observation. As shown by \citet{fs-eb-tilch-01}, under
certain circumstances the effect of an external parabolic potential is
also equivalent to an attractive force between the agents.  The emergence
of a vortex swarm can be seen as a \emph{dynamic compromise} between
three different requirements: active motion (to keep the agents moving),
asymmetric repulsive forces (e.g. to avoid collisions) and attractive
forces (either enforced by a local agent-agent attraction, or an external
potential). While the first two requirements alone would simply lead to a
dispersion, the latter one results in a compression (or confinement) of
the swarm. From a physics perspective the vortex swarm is a stable
attractor of the multi-agent dynamics; from an ``economics'' perspective
one could think of global utility maximization balancing out all
individual requirements (such as avoiding discomfort from both collisions
and separation).
  
A noticeable advantage of our model is that it does not break down for
small swarm sizes, so it can be used to simulate \emph{both} single
animals and swarms of low and of high density, whereas other models
mainly concentrate on the dynamics of reasonably sized swarms.  This
allows in particular to investigate the transition from the uncorrelated
(bi-directional) rotation of single agents to the correlated vortex
formation of the swarm, as shown in Fig. \ref{fig:phase_1}.
  
So far, we have shown a \emph{qualitative agreement} between our computer
simulations and some experimental observations in \emph{Daphnia}. A
quantitative verification, or even a prediction of \emph{Daphnia}
behavior under different real conditions is still missing. In this
respect, our model is not different from other theoretical models
proposed. A first step towards verification involves the experimental
determination of the parameters. In this paper, we can just propose some
ideas:

\begin{itemize}
\item Neglecting the jump-like motion of single \emph{Daphnia} and using
  the approximation of a (quasi-)stationary velocity, one should be able
  to estimate an average velocity of cycling \emph{Daphnia}. This can be
  directly related to the stationary velocity $v_{0}$, \eqn{v0}, that
  enters the equation of motion used in the model.
\item The parameters determining the avoidance behavior may be estimated
  by a direct comparison between the experimental observation, Fig.
  \ref{fig:avoidance_track} and the computer simulation, Fig.
  \ref{fig:avoidance_sim}. Obviously, the sole event shown in this paper,
  is not sufficient for that; so, we would expect further experimental
  investigations here.
\item Another way to (indirectly) estimate the parameters of the
  avoidance potential is via the local swarm density, or the (average)
  spatial extension of a swarm of given size. These are determined by our
  model parameters and could be possibly compared with experiments
  (varying both the strength of the light beam and the swarm size).
\item By varying the swarm size, one can also experimentally test the
  onset of the vortex swarming, and compare this to the respective
  computer simulations shown in Fig. \ref{fig:phase_1}. As mentioned, the
  transition range towards the vortex formation in the model strongly
  depends both on swarm size and on the parameters characterizing the
  avoidance potential. So, in addition to the observation of avoidance
  maneuvers of two animals, this yields a macroscopic verification for
  the parameters of the avoidance potential.
\end{itemize}
 
Eventually, we want to point out some situations where the model could
make predictions about \emph{Daphnia} behavior which may be tested
experimentally.  The model uses the assumption of an environmental
potential, that in the current investigation results in an attraction
toward the center, this way considering the influence of the vertical
light beam on the \emph{Daphnia}. The real influence could be tested by
producing \emph{Daphnia} mutants that are insensitive to light. Then we
expect no attractive force, and hence no vortex swarming, in agreement
with the explanations above. Also, one could think of \emph{two
  different} vertical light beams in the water tank, at a certain
distance. Starting from a homogeneous spatial distribution of agents, the
model would predict the occurrence of two different vortex swarms around
the two centers, each probably having its own rotational direction, as
long as the distance between the light beams is large enough. For smaller
distances (where the critical distance may be also a function of the
agent density in the system), we would expect from the model
interferences at the boarder between the two rotating swarms, which would
lead to additional couplings and thus maybe to a synchronization of the
rotational directions.  The situation of the two separated vertical light
beams would also allow us to test whether the attraction of the
\emph{Daphnia} is limited by some \emph{maximum local density} of the
swarm, as suggested by our model. In this case, we would find in the
experiments two distinct swarms, whereas without saturation effects, one
could possibly find one swarm only, with a much higher density.

A remaining question is weather the model in the current form is also
applicable to other species. Vortex swarming, as we have pointed out, is
a widely spread phenomenon observed also in fish, or bacteria.  We argue
that the principle (qualitative) features of vortex swarming are covered
by our model, as long as (local or global) attraction and asymmetric
repulsion (e.g. via avoidance maneuvers) play a considerable role. This
may hold for fish, but probably not for gliding bacteria like
\emph{Paenibacillus}, which also show vortex formation
\citep{benjacob-03}. In the latter case, adhesion forces may play a much
more important role as one can also deduce by looking at the sharp
external boundaries of the bacterial swarm. However, it would be still
possible to adapt the model of Brownian agents also for this situation by
including other terms of local interaction.

\subsection*{Acknowledgment}
The authors thank Anke Ordemann for discussion and for providing data of
the \textit{Daphnia}-\textit{Daphnia} avoidance maneuver investigated in
Sect.~\ref{sec:gath-empir-evid}.

\bibliography{daphnia-fs-web}

\end{document}